\documentclass[aps,twocolumn,showpacs]{revtex4}
\usepackage{graphicx}
\usepackage{amsmath}
\usepackage{amssymb}
\usepackage{float}
\begin{document}

\title{Geometry-induced chaos and scaling near two singular limits of an eye-shaped billiard}

\author{Anne K\'etri P. da Fonseca$^{1,2}$, Edson D.\ Leonel$^2$, Diego F. M. Oliveira$^1$}

\affiliation{$^1$School of Electrical Engineering and Computer Science, University of North Dakota, USA\\
$^2$Departamento de F\'isica, Unesp - Universidade Estadual Paulista - 
Av.24A. 1515, 13506-900, Rio Claro, SP, Brazil}

\date{\today}

\begin{abstract}
We investigate how boundary geometry controls the coexistence of
regular and chaotic motion in a static, eye-shaped billiard.
A single control parameter continuously deforms the smooth,
strictly convex boundary between two singular geometric limits.
Across the sampled parameter range, chaotic layers coexist with
regular islands and invariant curves that restrict phase-space
transport. We derive analytical conditions for the existence of
diagonal period-two orbits and determine the stability transitions
of the axial period-two trajectories. We also establish the
existence of invariant caustics near the boundary for every finite
positive value of the control parameter.
Numerical simulations reveal a nonmonotonic dependence of the
average maximal Lyapunov exponent on the control parameter, with
the strongest sampled instability occurring at an intermediate
geometry. The mean squared-angle envelope width associated with
the sampled chaotic component decreases toward both singular
limits, following power laws with fitted exponents of opposite
signs and magnitudes close to two thirds. We further characterize
a geometric response function and discuss the resulting analogy
with continuous critical behavior. Our findings show how
geometric deformation governs orbital stability, barriers to
chaotic transport, and scaling properties in conservative
billiard dynamics.
\end{abstract}

\maketitle


\section{Introduction}
\label{xsec1}
Collision models are paradigmatic examples of dynamical systems in which conceptually simple rules give rise to complex behavior. The Fermi-Ulam \cite{fermi,ulam1961} and bouncer \cite{pustylnikov1983,pustylnikov1987,pustylnikov1995} models, for example, describe the one-dimensional motion of a particle undergoing collisions with moving boundaries. Specifically, the Fermi-Ulam model consists of a classical particle bouncing between a fixed and an oscillating wall, while the bouncer model features a particle falling under a constant gravitational field onto a periodically moving platform. Despite their underlying simplicity, both have become seminal systems for understanding chaotic transport \cite{lichtenberg1992,leonel2004,altmann2013} and Fermi acceleration \cite{lieberman1972,gelfreich2008}, contributing significantly to the development of nonlinear dynamical systems theory \cite{schmelcher2006,lenz2008}. Energy diffusion in collision mappings has also been investigated through biased-random-walk descriptions linking asymmetry in velocity distributions to unlimited energy growth \cite{oliveira2015symmetry}. Quantum versions of these models have also been investigated \cite{quantum1,quantum2}.

The mathematical study of billiard dynamics dates back at least to Birkhoff's work in the 1920s \cite{ref01} on the motion of a free particle confined inside a closed domain. In general, a billiard is defined by a connected region $Q\subset\mathbb{R}^D$, whose boundary $\partial Q$ is a $(D-1)$-dimensional hypersurface, possibly piecewise smooth, embedded in $\mathbb{R}^D$. Inside the billiard, a point particle of mass $m$ (commonly set to unity) moves freely along a straight line until it encounters the boundary, where it undergoes specular reflection \cite{chernov2006chaotic}. Depending on the boundary geometry and control parameters, billiards can exhibit regular, ergodic, or mixed dynamics. In mixed systems, the behavior of an individual trajectory also depends on its initial condition. Integrability is associated with additional conserved quantities, such as angular momentum in the circular billiard. The Bunimovich stadium \cite{ref2,ref3} and Sinai billiard \cite{ref1} are standard examples of ergodic billiards. In these systems, almost every initial condition generates a trajectory whose long-time averages agree with phase-space averages with respect to the invariant measure. Ergodicity does not, however, require every periodic orbit to be unstable. Many other billiards exhibit mixed phase-space structures \cite{Ref6,Ref7,Ref8,Ref9,Ref10a,Ref10b,xx1,hermes2022break,miranda2021short}, controlled by parameters with different physical meanings. Their phase spaces can contain invariant spanning curves (sections of invariant tori), regular islands surrounding elliptic periodic orbits, and chaotic seas. Such coexistence is common in nonintegrable Hamiltonian systems \cite{add2} and is also observed in one-dimensional time-dependent potentials \cite{Ref11,Ref13,graciano2022multiple}, rippled channels \cite{Ref17}, and models of tokamak plasmas \cite{Ref192,Ref21}.

Depending on the billiard geometry, variation of a control parameter can lead from integrable to mixed or chaotic dynamics. A standard framework for studying the loss of integrability is the near-integrable Hamiltonian
$H(I_1,\theta_1,I_2,\theta_2)=H_0(I_1,I_2)+\varepsilon H_1(I_1,\theta_1,I_2,\theta_2)$.
Related transitions also arise in parameter-dependent billiard maps \cite{leonel2020characterization,11nova,oliveira2013some}. A common example is the oval billiard, where $\varepsilon=0$ recovers the fully integrable circle and $\varepsilon\ne0$ deforms the boundary, producing mixed dynamics. Further deformation can destroy regular structures, depending on the boundary family and its smoothness and convexity \cite{oliveira2010suppressing}. Near integrability-breaking transitions, scaling laws can emerge and be characterized by critical exponents \cite{Ref11}. Scaling descriptions have also been developed for the hybrid Fermi-Ulam-bouncer model, where invariant spanning curves bound a chaotic region whose velocity fluctuations exhibit parameter-dependent scaling \cite{oliveira2009hybrid}. In many systems, these laws describe diffusion and transport, providing a framework for comparing dynamical behavior across different physical settings.

We consider a static eye-shaped billiard whose boundary is parameterized in Cartesian coordinates as $X(\theta)=\cos\theta$ and $Y(\theta)=\varepsilon\sin\theta(1+\sin^2\theta)$. Its aspect ratio is controlled by $\varepsilon$, which changes the relative extent of regular and chaotic regions in phase space. We characterize the periodic orbits, the surviving invariant curves, the maximal Lyapunov exponent, and a geometric measure of the chaotic envelope. The Jacobian written in noncanonical coordinates provides an additional diagnostic of the collision map, while the natural billiard measure remains invariant.

The limit $\varepsilon\to0$ produces a horizontal line segment, whereas $\varepsilon\to\infty$ produces a vertical segment after uniform spatial rescaling by $1/\varepsilon$. Neither limiting object defines a nondegenerate two-dimensional billiard. We therefore study the approach to these \emph{singular geometric limits} from within the family $0<\varepsilon<\infty$. The question is how the sampled chaotic envelope shrinks and the surrounding invariant structures reorganize as each limiting geometry is approached. The resulting power laws motivate a comparison with continuous-transition phenomenology, although no thermodynamic free energy is defined for the present system. We distinguish the exponent of the measured mean squared-angle envelope width from that of its square root and define a response function with respect to an explicitly chosen geometric distance from each limit.

This paper is organized as follows. Section~\ref{xsec2} defines the model and its collision map. Section~\ref{xsec3} discusses the noncanonical Jacobian, analytically accessible periodic orbits, Lyapunov exponents, and invariant caustics. Section~\ref{xsec4} defines the chaotic-envelope observable and examines its asymptotic scaling. Section~\ref{xsec5} discusses a geometric response function and the status of the transition analogy. Section~\ref{xsec6} summarizes our conclusions.

\section{The billiard model}
\label{xsec2}
\begin{figure}[h]
    \centering
    \includegraphics[width=\linewidth]{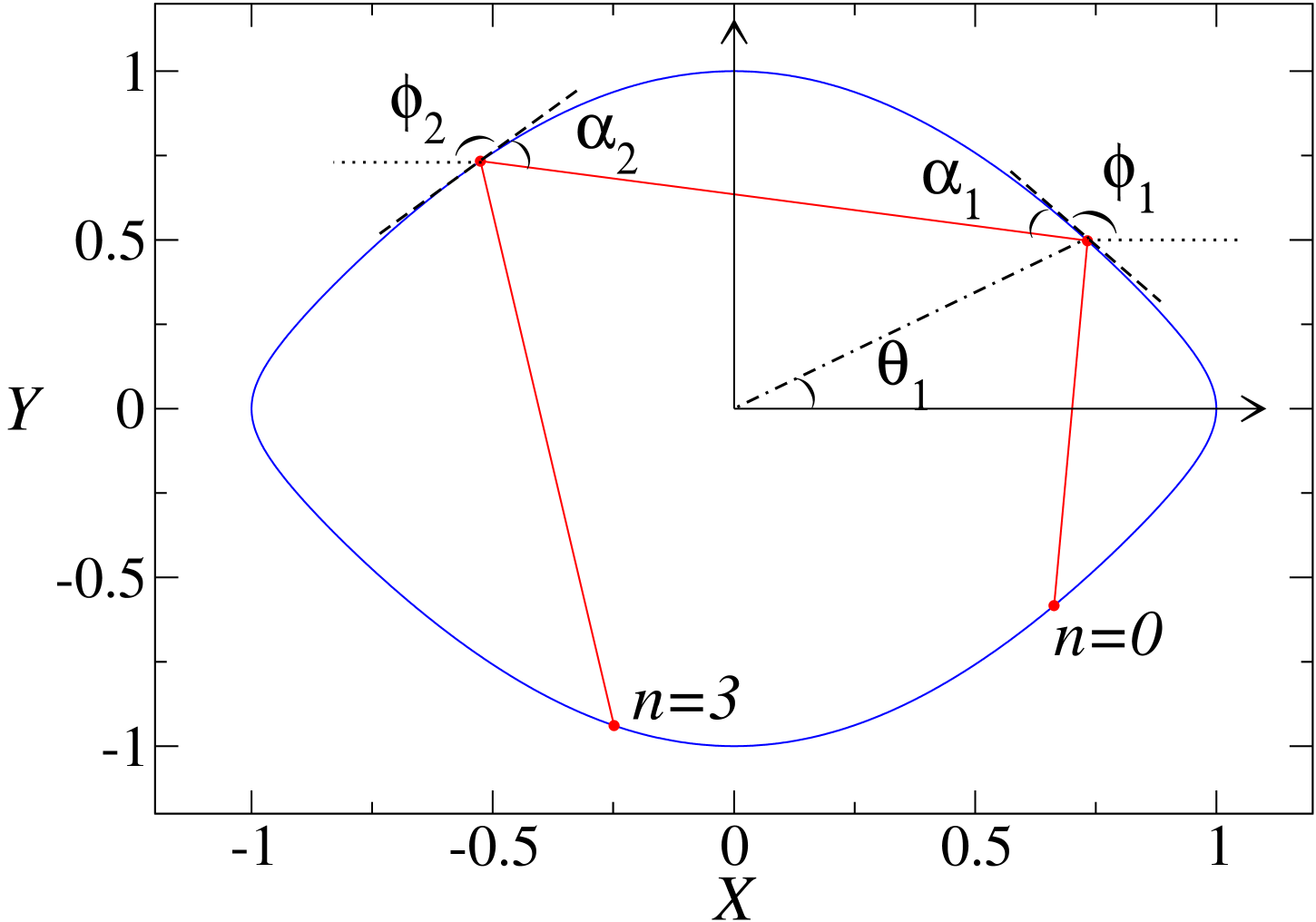}
    \caption{Boundary of the eye billiard (in blue) with its corresponding geometric parameters ($\varepsilon=0.5$). A representative particle trajectory is shown in red for the first few collisions.}
    \label{fig1}
\end{figure}

We now construct a nonlinear map that describes the dynamics of the model.
The model consists of a classical particle of mass $m$ confined within
the billiard and moving freely in the absence of external fields.
When the particle hits the boundary, it undergoes specular reflection
without any loss of energy. The dynamics is described by a
two-dimensional nonlinear map
$T(\theta_n,\alpha_n)=(\theta_{n+1},\alpha_{n+1})$, where
$\theta_n$ denotes the boundary parameter at the collision point.
The variable $\alpha_n$ is the angle that the outgoing trajectory
makes with the tangent vector to the boundary, as illustrated in
Fig.~\ref{fig1}. The index $n$ denotes the $n$th collision.
The boundary is defined parametrically, with $\theta$ serving as
a geometric parameter rather than the polar angle.
The Cartesian coordinates $[X(\theta_n),Y(\theta_n)]$ are given by:
\begin{equation}
\begin{cases}
X(\theta_n)=\cos(\theta_n), \\
Y(\theta_n)=\varepsilon \sin(\theta_n)[1+ \sin^2(\theta_n)].
\label{eq12}
\end{cases}
\end{equation}
Fig.~\ref{fig1} shows the geometry of the boundary for $\varepsilon=0.5$ as well as the trajectory of a particle for the first few collisions.  

Given an initial condition $(\theta_n,\alpha_n)$, the tangent direction at
$[X(\theta_n),Y(\theta_n)]$, taken modulo $\pi$ consistently with the
unoriented tangent line, is written as
\begin{eqnarray}
\phi_n=\arctan\left[ \frac{Y'(\theta_n)}{X'(\theta_n)} \right]~,
\label{eq3}
\end{eqnarray}
where $X'$ and $Y'$ denote derivatives with respect to $\theta$. In a numerical implementation, the tangent direction should be evaluated with the two-argument function $\operatorname{atan2}(Y',X')$ and the angle conventions handled modulo $\pi$; this avoids quadrant ambiguities in the single-argument arctangent. The derivatives are
\begin{equation}
\begin{cases}
X'(\theta_n)=-\sin(\theta_n), \\
Y'(\theta_n)=\varepsilon \cos(\theta_n)[1+3\sin^2(\theta_n)].
\end{cases}
\label{eq13}
\end{equation}

Since the particle travels with constant velocity along a straight line until it reaches the boundary, we must solve:
\begin{equation}
Y(\theta_{n+1})-Y(\theta_n)=\tan(\alpha_n+\phi_n)[X(\theta_{n+1})-X(\theta_n)],
\label{eq5}
\end{equation}
to obtain the next collision with the boundary. The physical solution is the first nontrivial intersection along the outgoing ray, excluding the current collision point. The slope form in Eq.~(\ref{eq5}) is equivalent, where defined, to a cross-product equation for the ray and chord; the latter avoids numerical difficulties when the trajectory is vertical. A numerical solver must select the forward intersection. Since $X(\theta)=\cos\theta$, a recovered Cartesian collision point can be mapped back to the geometric parameter via $\theta=\arccos X$, with $\theta\mapsto2\pi-\theta$ when $Y<0$. Unlike the polar angle $\arctan(Y/X)$, this reconstruction respects the parametric boundary.

The new angle that the trajectory makes with the tangent
at $\theta_{n+1}$ is obtained from geometrical considerations, as can be
seen in Fig. \ref{fig1}. It is given by:
\begin{equation}
\alpha_{n+1}=\phi_{n+1}-(\alpha_n+\phi_n).
\label{eq6}
\end{equation}

Thus, the complete map that
describes the dynamics of the model is given by:
\begin{equation}
T:\begin{cases}
F(\theta_{n+1})=Y(\theta_{n+1})-Y(\theta_{n})-\\ \ \ \tan(\alpha_{n}+\phi_{n})[X(\theta_{n+1})-X(\theta_{n})] = 0,~~\\
\alpha_{n+1}=\phi_{n+1}-(\alpha_{n}+\phi_{n})
\end{cases} 
\label{eq7}
\end{equation}
where $\theta_{n+1}$ is obtained from the next physical root of $F(\theta_{n+1})=0$ and $\phi_{n+1}$ is the tangent direction at the new collision. Angular variables are reduced to their physical branches after reflection.

\begin{figure*}[htbp]
    \centering
    \begin{minipage}[t]{0.32\linewidth}
        \raggedright (a)\\
        \centering
        \includegraphics[width=\textwidth]{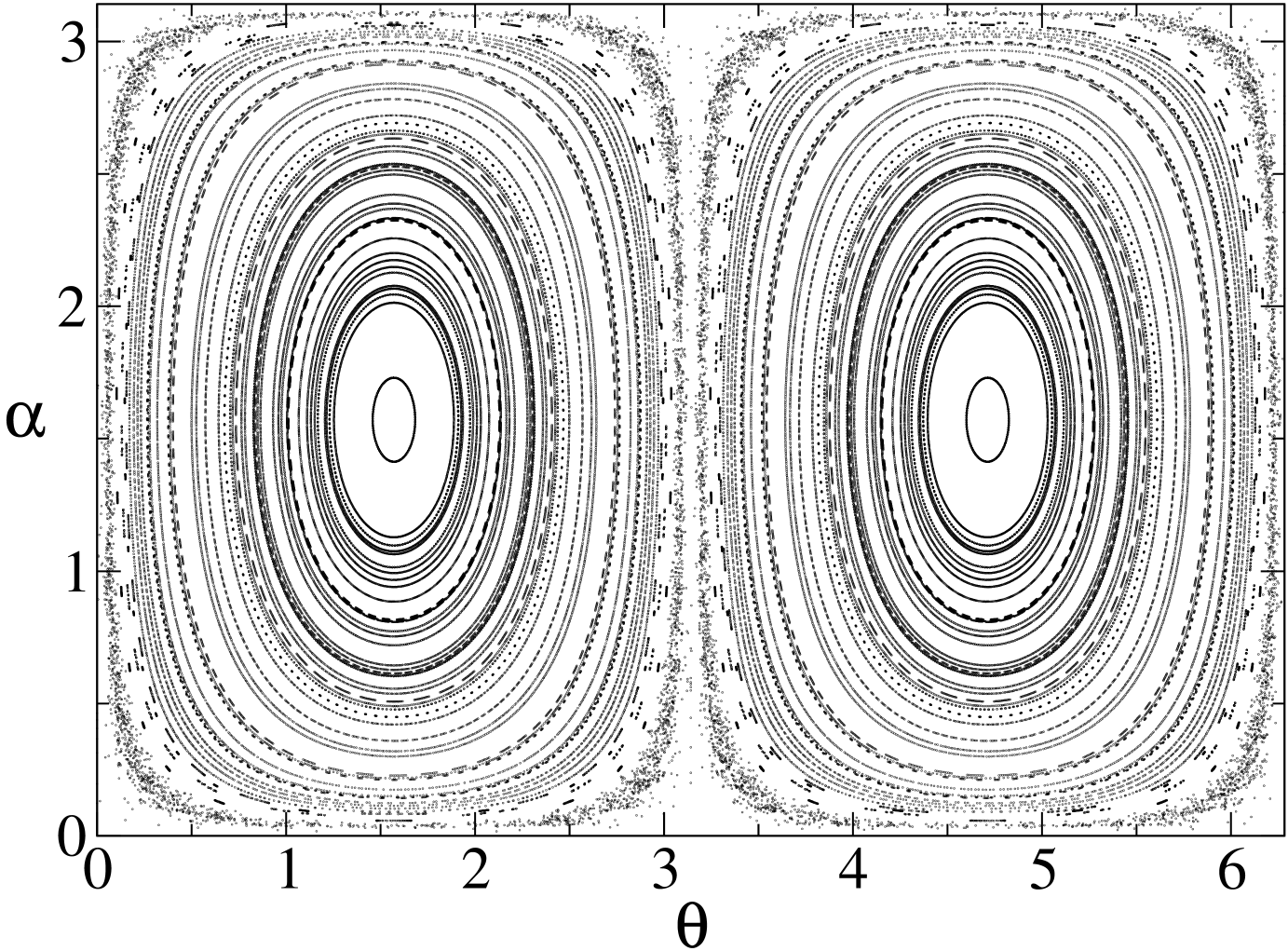}
    \end{minipage}%
    \hfill
    \begin{minipage}[t]{0.32\linewidth}
        \raggedright (b) \\
        \centering 
        \includegraphics[width=\textwidth]{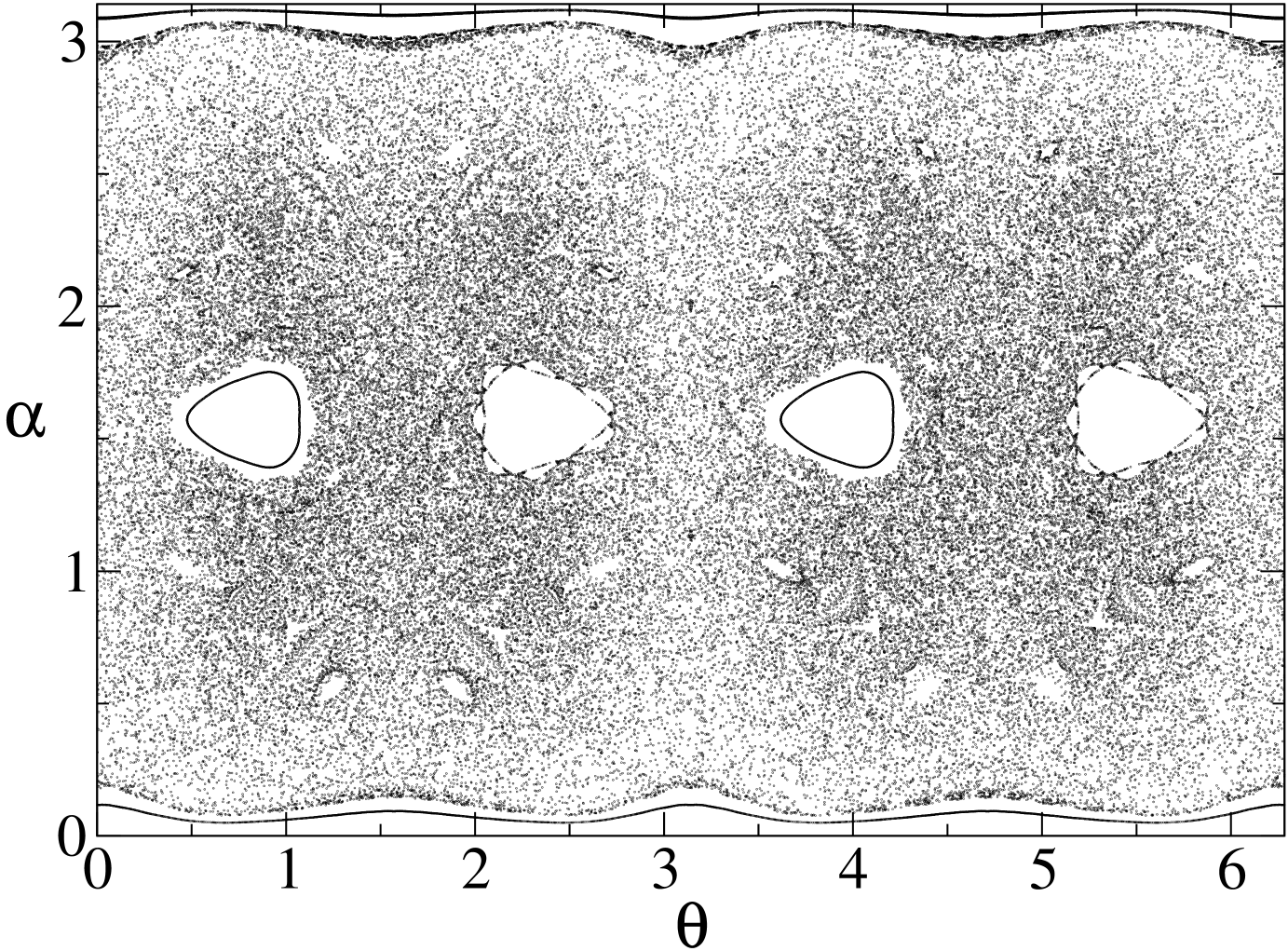}
    \end{minipage}%
    \hfill
    \begin{minipage}[t]{0.32\linewidth}
        \raggedright (c) \\
        \centering 
        \includegraphics[width=\textwidth]{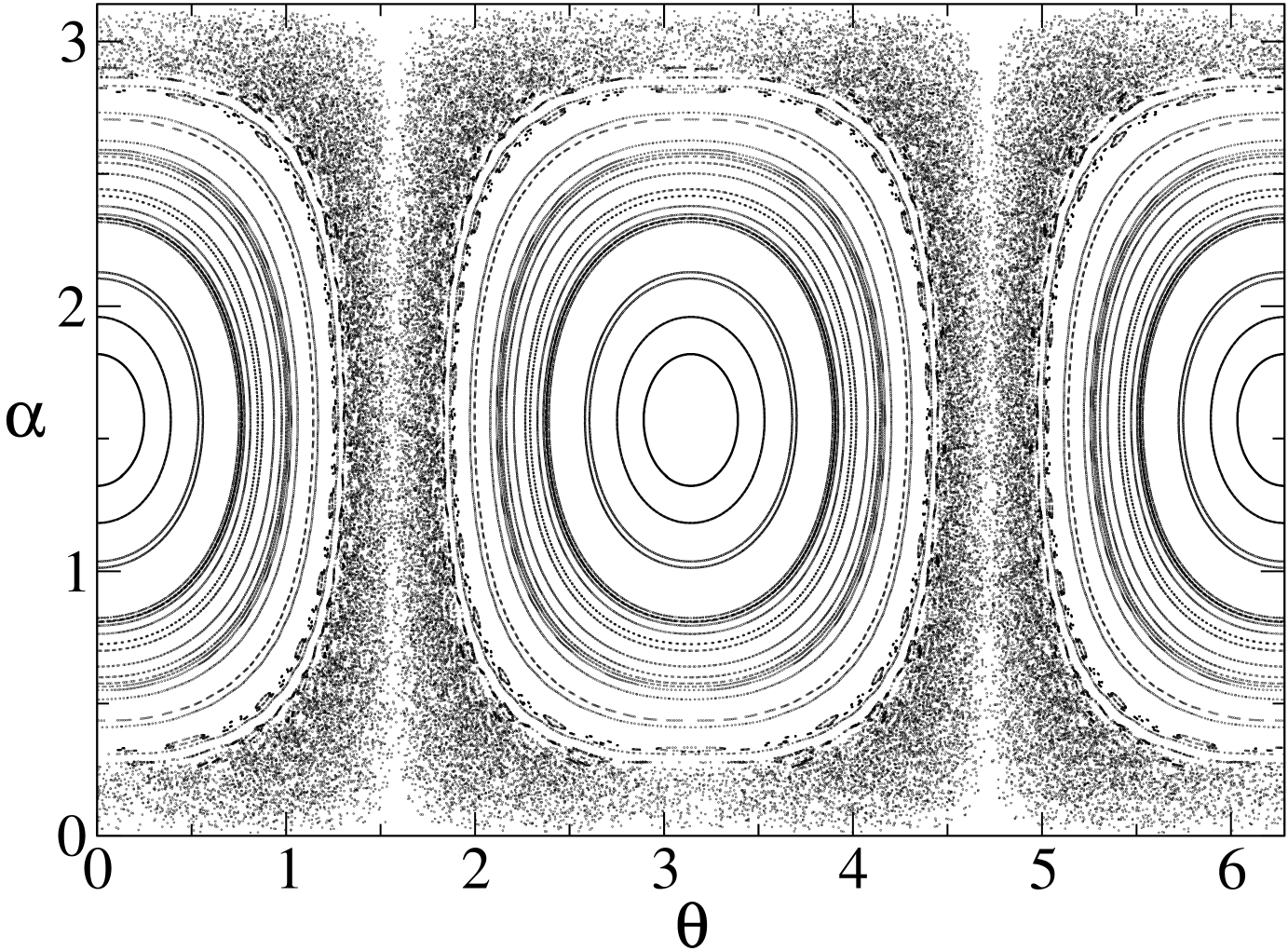}
    \end{minipage}%

    \vspace{0.3cm} %

    \begin{minipage}[t]{0.32\linewidth}
        \raggedright (d) \\
        \centering 
        \includegraphics[width=\textwidth]{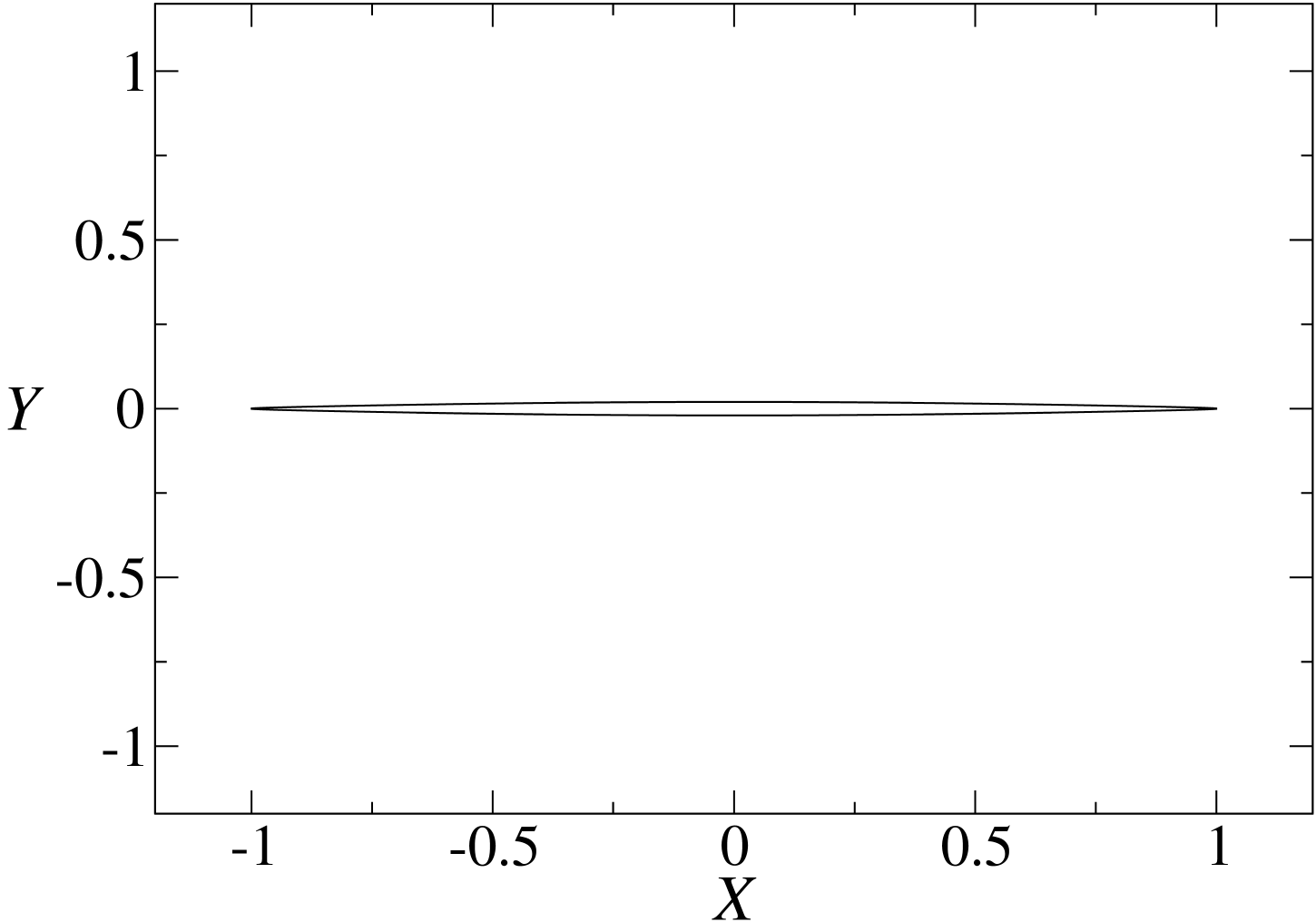}
    \end{minipage}%
    \hfill
    \begin{minipage}[t]{0.32\linewidth}
       \raggedright (e) \\
        \centering 
        \includegraphics[width=\textwidth]{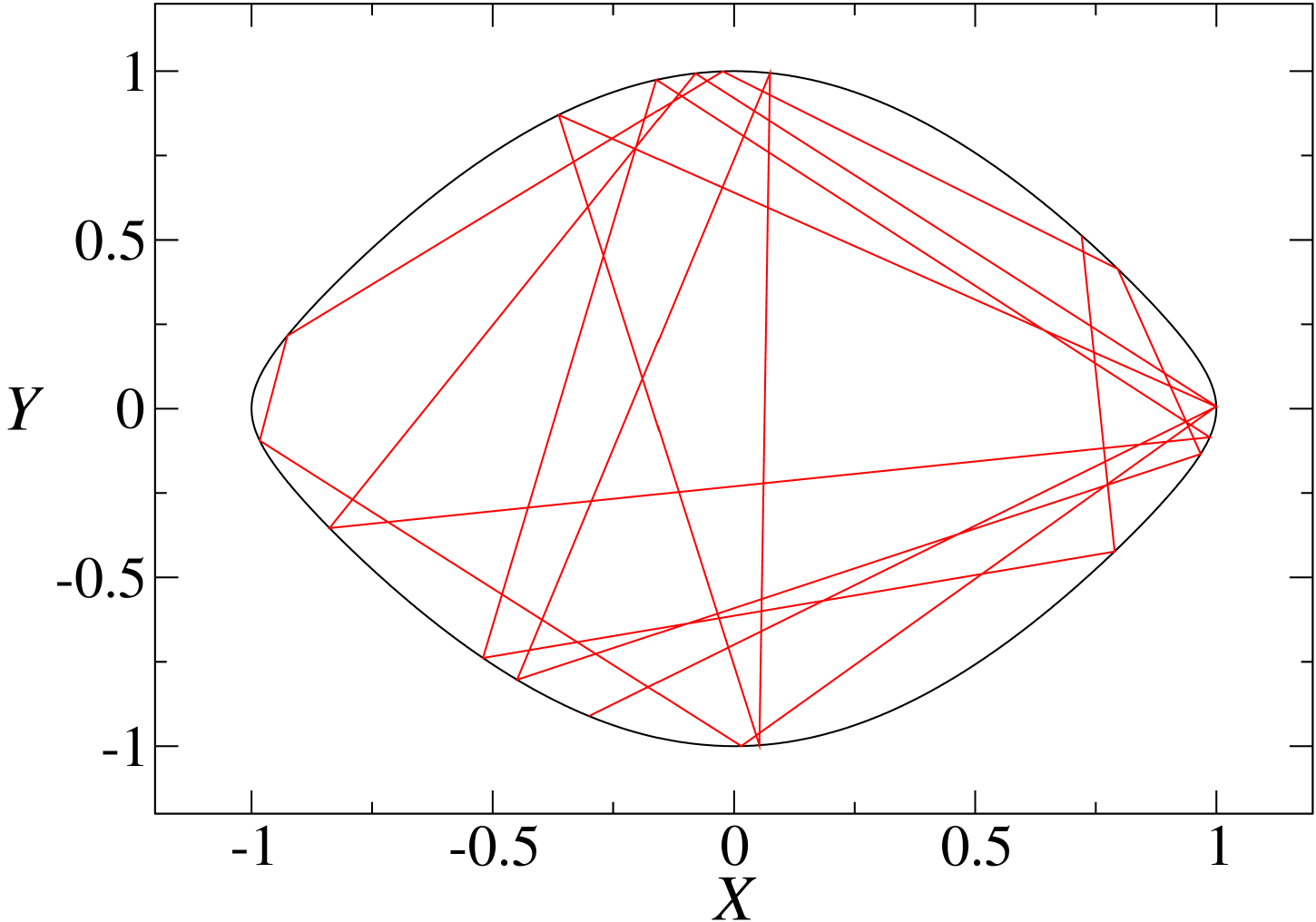}
    \end{minipage}%
    \hfill
    \begin{minipage}[t]{0.32\linewidth}
        \raggedright (f) \\
        \centering 
        \includegraphics[width=\textwidth]{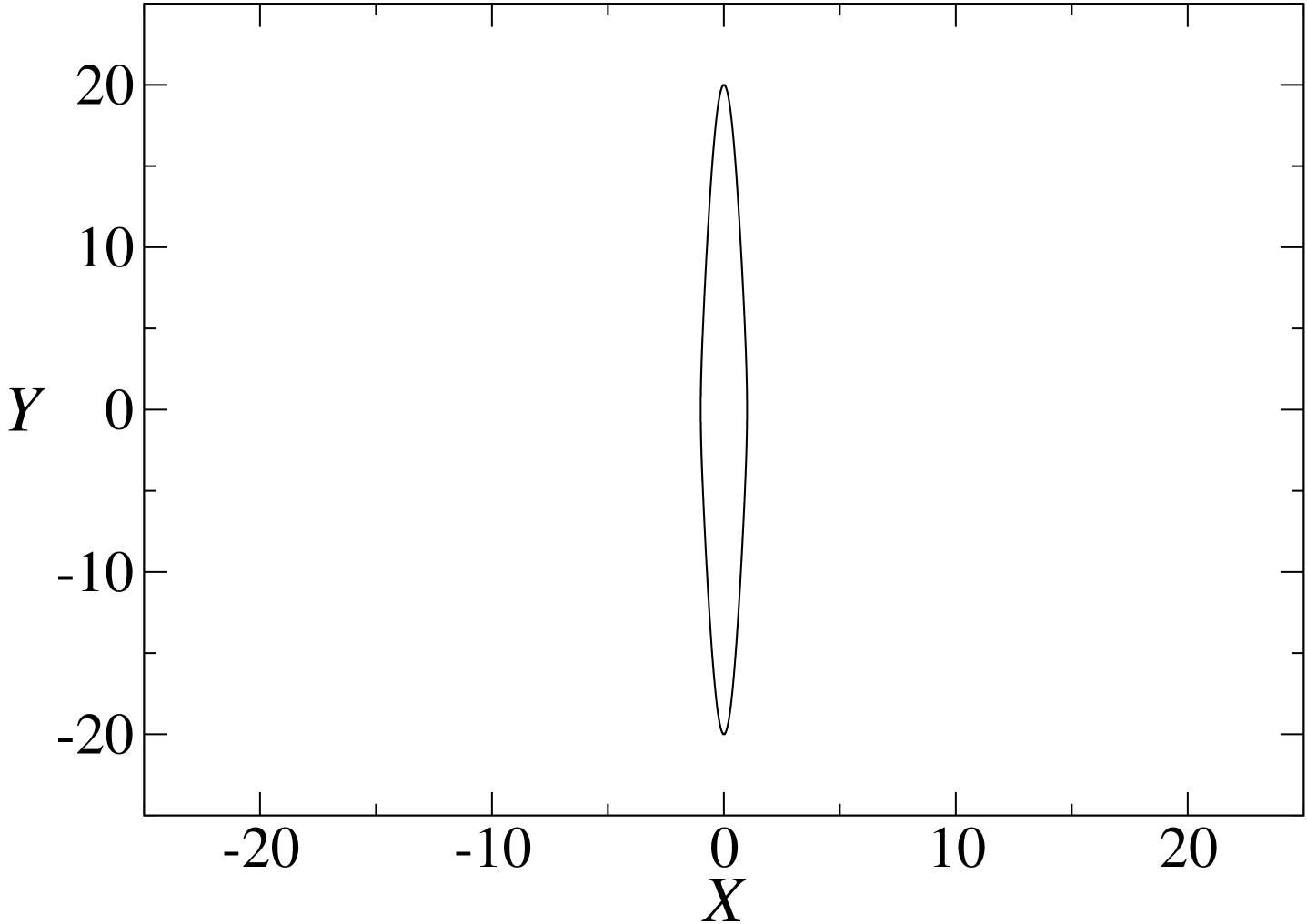}
    \end{minipage}%

    \vspace{0.3cm}

    \caption{Phase spaces and associated shapes of the boundaries for different combinations of control parameters: $\varepsilon=0.01$ in panels (a) and (d),  $\varepsilon=0.5$ in panels (b) and (e) and $\varepsilon=0.10$ in panels (c) and (f).}
    \label{fig002}
\end{figure*}

To illustrate the behavior of the phase space, we iterate Eq. (\ref{eq7}) using three different values of $\varepsilon$: $\varepsilon=0.01$, $\varepsilon=0.5$, and $\varepsilon=10$, as presented in Figure \ref{fig002}.

The phase space displays a rich structure containing invariant spanning curves (sections of invariant tori), Kolmogorov-Arnold-Moser (KAM) islands, and chaotic seas \cite{lichtenberg1992}. The family approaches two singular geometric limits as $\varepsilon\to0$ and $\varepsilon\to\infty$. In the first limit, the boundary collapses onto a horizontal line segment; in the second, it approaches a vertical line segment after uniform spatial rescaling by $1/\varepsilon$. Neither limit defines a nondegenerate two-dimensional billiard.
Near either singular limit, the geometry becomes strongly flattened or elongated, and repeated nearly normal collisions resemble motion between opposing static walls. The phase spaces display extensive regular islands, while rotational invariant curves survive close to grazing incidence ($\alpha\to0$ and $\alpha\to\pi$). These whispering-gallery trajectories remain possible for every finite positive $\varepsilon$, as shown below.

\begin{figure}[h!]
\centering
    \raggedright (a) \\ \centerline{\includegraphics[width=0.8\linewidth]{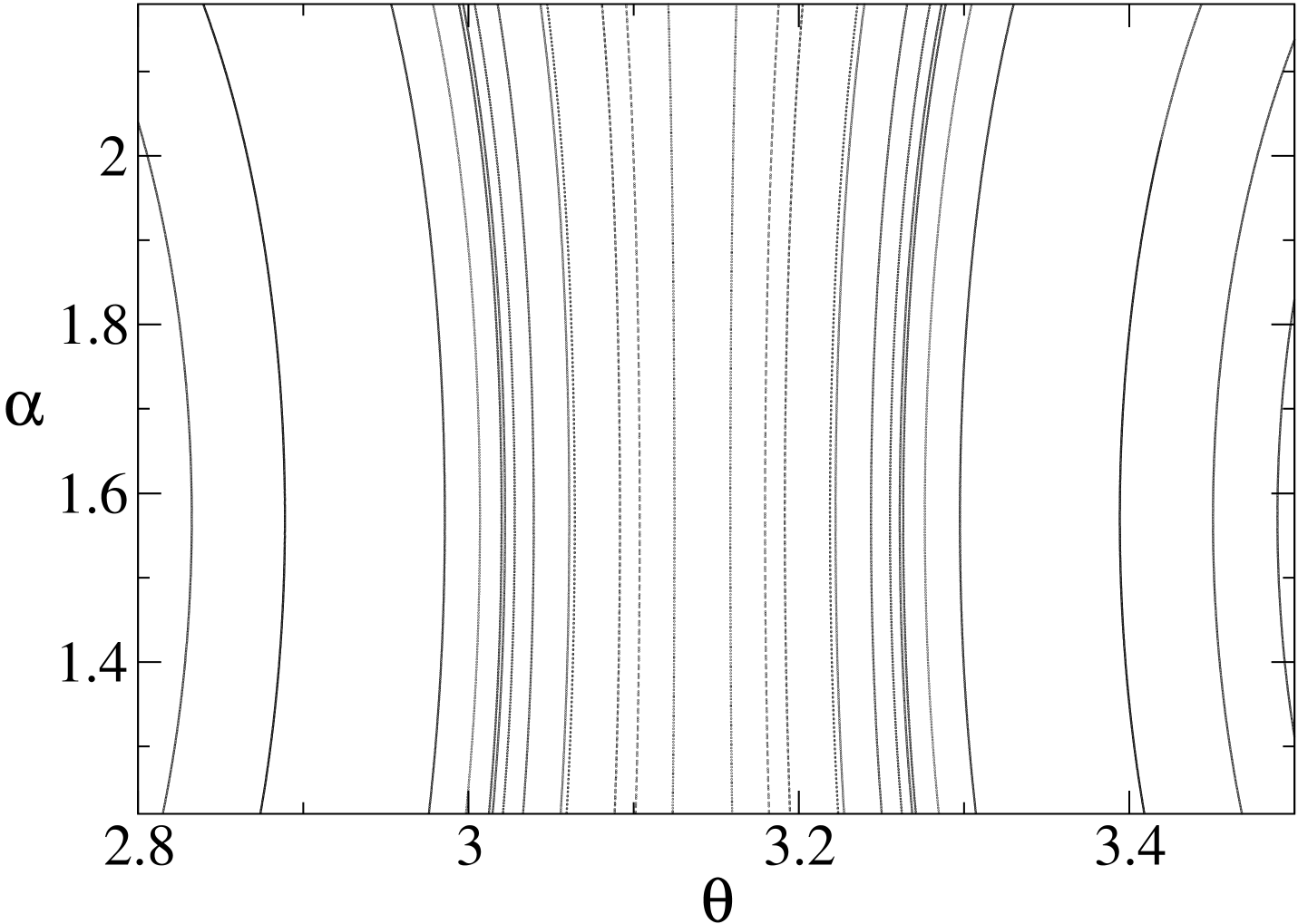}}
    \raggedright (b) \\
    \centerline{\includegraphics[width=0.8\linewidth]{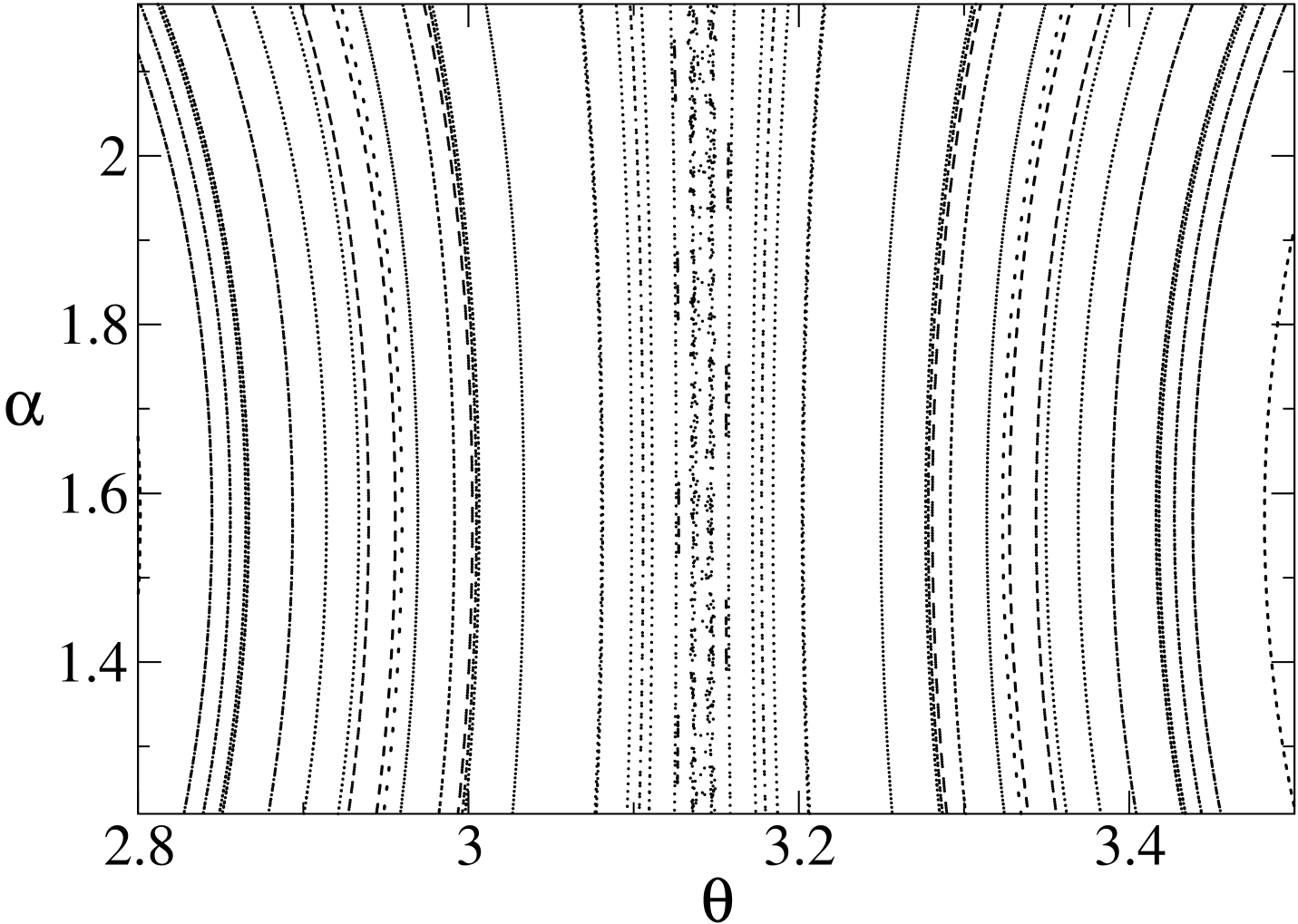}}    

    \raggedright (c) \\
    \centerline{\includegraphics[width=0.8\linewidth]{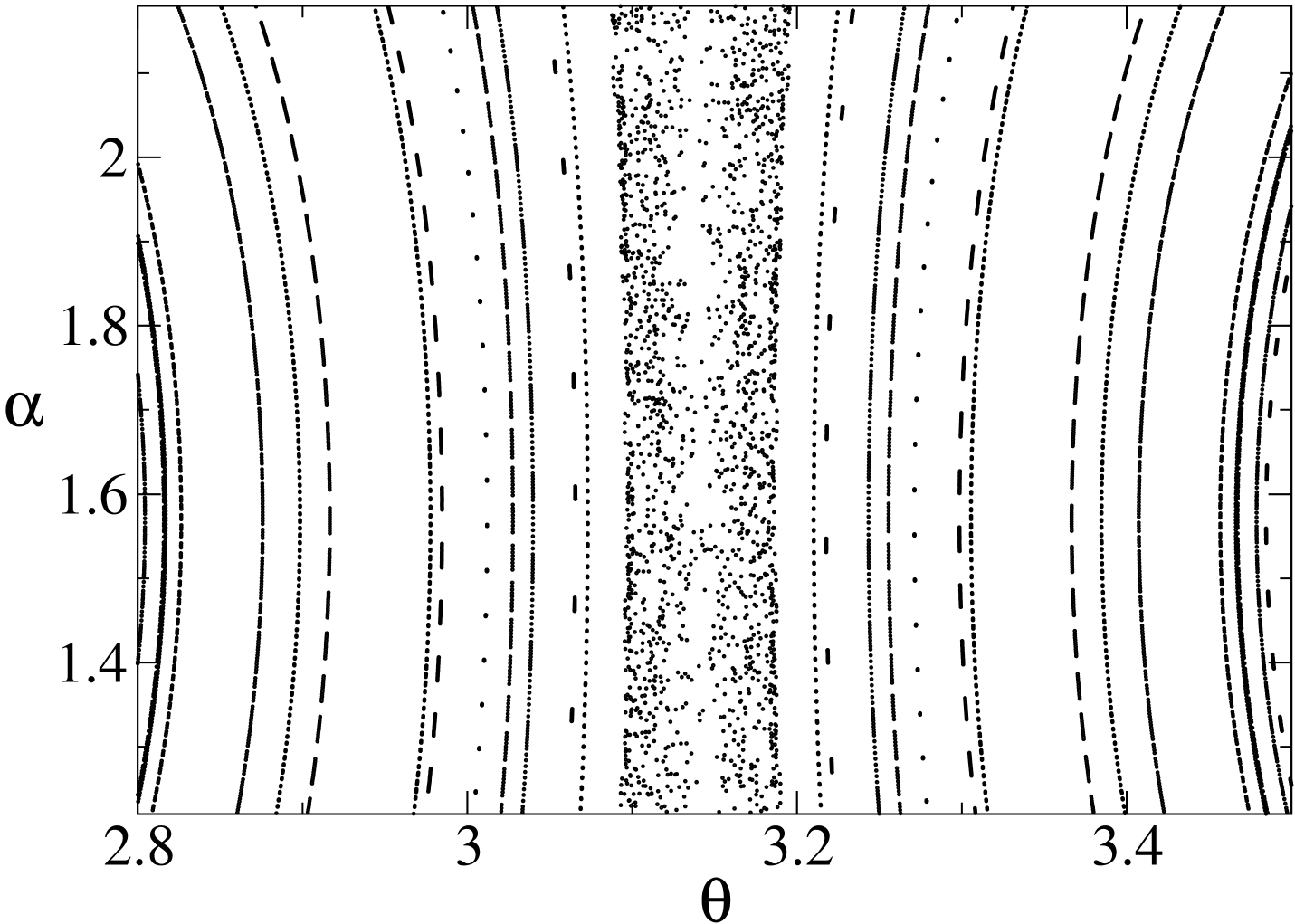}}  
\caption{Magnification of the center region of the phase space for different
control parameters: (a)$\varepsilon=5\times10^{-3}$, (b)$10^{-3}$ and (c)$10^{-2}$}
\label{fig003}
\end{figure}

The magnification of the region around $(\pi,\pi/2)$ presented in Fig.~\ref{fig003} reveals that a chaotic layer persists even for small values of $\varepsilon$, between the invariant curves and the regular islands. Its sampled envelope narrows as $\varepsilon\to0$, with analogous behavior observed as $\varepsilon\to\infty$.

As observed in other billiards \cite{separatriz}, the chaotic layer lies in the region separating regular islands associated with libration from invariant curves corresponding to rotation. However, a distinct feature of the eye billiard is that the chaotic layer does not simply become confined around $\alpha=\pi/2$ as a singular limit is approached. Instead, the numerical portraits indicate that portions of the layer extend toward the rotational invariant curves near $\alpha=0$ and $\alpha=\pi$. The behavior of these invariant curves and the associated scaling properties of the chaotic envelope are examined in the following sections.

\section{Numerical results}
\label{xsec3}
\subsection{Jacobian Determinant and Fixed Points}
To characterize how the billiard map is represented in the coordinates $(\theta,\alpha)$, we consider its Jacobian matrix. A determinant different from unity describes deformation of coordinate-area elements; it does not imply physical dissipation in this conservative model. Canonical billiard coordinates are the Birkhoff pair $(s,p)$, where $s$ is the boundary arc length and $p=\cos\alpha$ when $\alpha$ is measured from the tangent \cite{Arnold,chernov2006chaotic}. We use $(\theta,\alpha)$ for convenience, noting that the invariant measure is $d\mu=s'(\theta)\sin\alpha\,d\theta\,d\alpha$, up to normalization,  with $s'(\theta)=\sqrt{X'(\theta)^2+Y'(\theta)^2}$.

From the mapping we can write the Jacobian matrix for $(\theta_{n+1},\alpha_{n+1})$ as:
\begin{equation}
    J=\begin{pmatrix}
j_{11} & j_{12} \\
j_{21} & j_{22} 
\end{pmatrix}	
\end{equation}
where
\begin{gather*}
    j_{11}=\frac{\partial\theta_{n+1}}{\partial\theta_n} \ \ \ \ \ \ \ j_{12}=\frac{\partial\theta_{n+1}}{\partial\alpha_n}\\
     j_{21}=\frac{\partial\alpha_{n+1}}{\partial\theta_n}  \ \ \ \ \ \ \ j_{22}=\frac{\partial\alpha_{n+1}}{\partial\alpha_n}
\end{gather*}
Once all the terms are properly defined we can obtain the Jacobian determinant by the product
\begin{equation}
    \det J=j_{11}j_{22}- j_{12}j_{21}=-j_{11}+j_{12}\frac{\partial\phi_n}{\partial \theta_n};
\end{equation}
which can be simplified to
\begin{equation}
    \det J= -\frac{Y'(\theta_n) - X'(\theta_n)\tan(\phi_n + \alpha_n)}{Y'(\theta_{n+1}) - X'(\theta_{n+1})\tan(\alpha_n+\phi_n)}
\end{equation}
Additionally, defining the differential arc length function $s'(\theta)=\sqrt{X'(\theta)^2+Y'(\theta)^2}$ and doing the proper simplifications, this  equation becomes
\begin{equation}
    \det J= \frac{\sqrt{X'(\theta_{n})^2+Y'(\theta_{n})^2}\sin \alpha_{n}}{\sqrt{X'(\theta_{n+1})^2+Y'(\theta_{n+1})^2}\sin \alpha_{n+1}}
\end{equation}
where $s'(\theta)d\theta=ds$. The numerator and denominator follow directly from the identities $Y'-X'\tan(\phi+\alpha)=-s'\sin\alpha/\cos(\phi+\alpha)$ at the departing collision and $Y'-X'\tan(\phi_n+\alpha_n)=s'\sin\alpha_{n+1}/\cos(\phi_n+\alpha_n)$ at arrival, with consistent angular branches. Equation~(10) is therefore the ratio of the invariant-measure densities at successive collisions, as required by the area-preserving canonical map.

\begin{figure}[h!]
\centering
    \raggedright (a)\vspace{-15pt}
    \centerline{\includegraphics[width=0.8\linewidth]{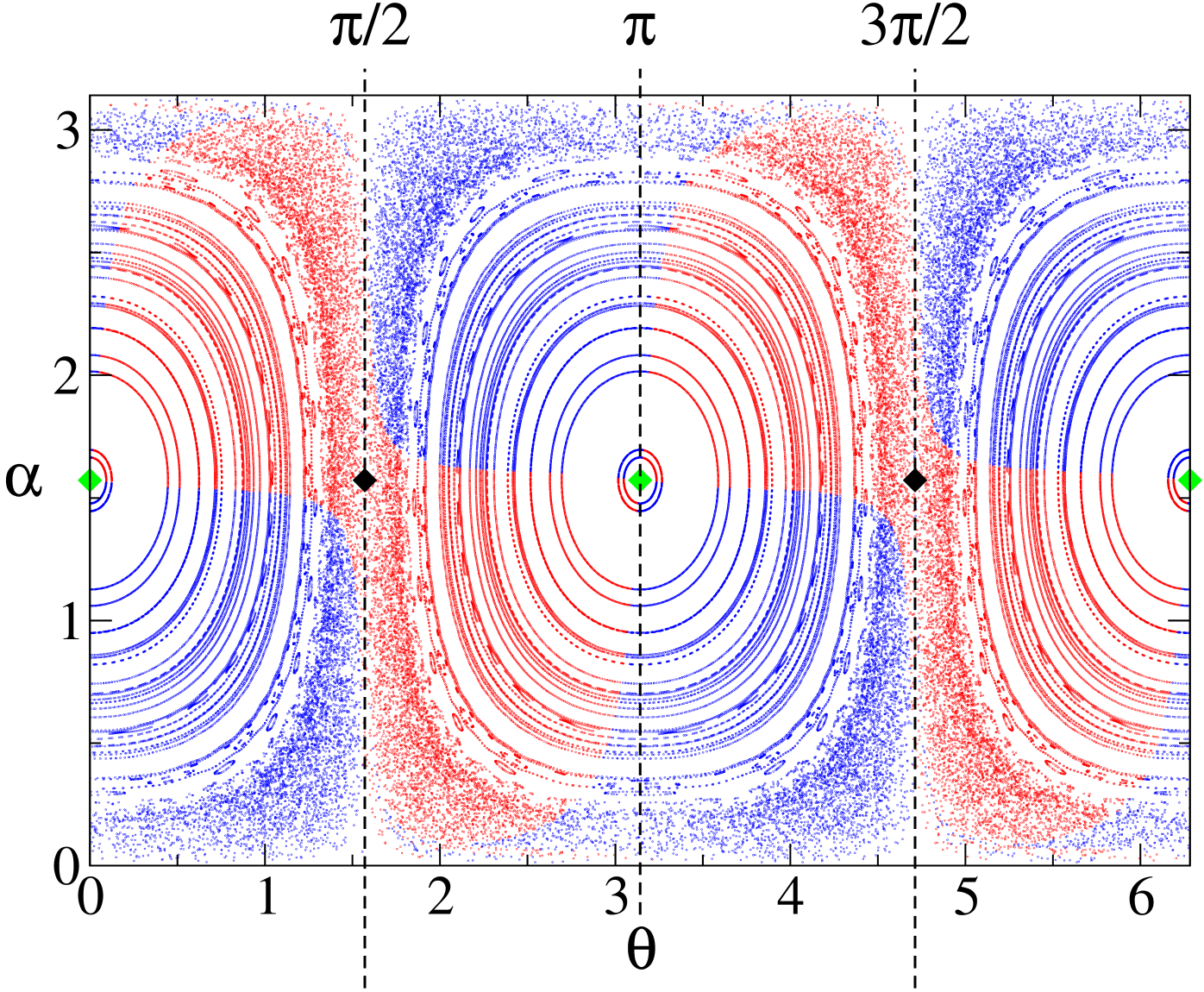}}
    \raggedright (b)\vspace{-15pt}
    \centerline{\includegraphics[width=0.8\linewidth]{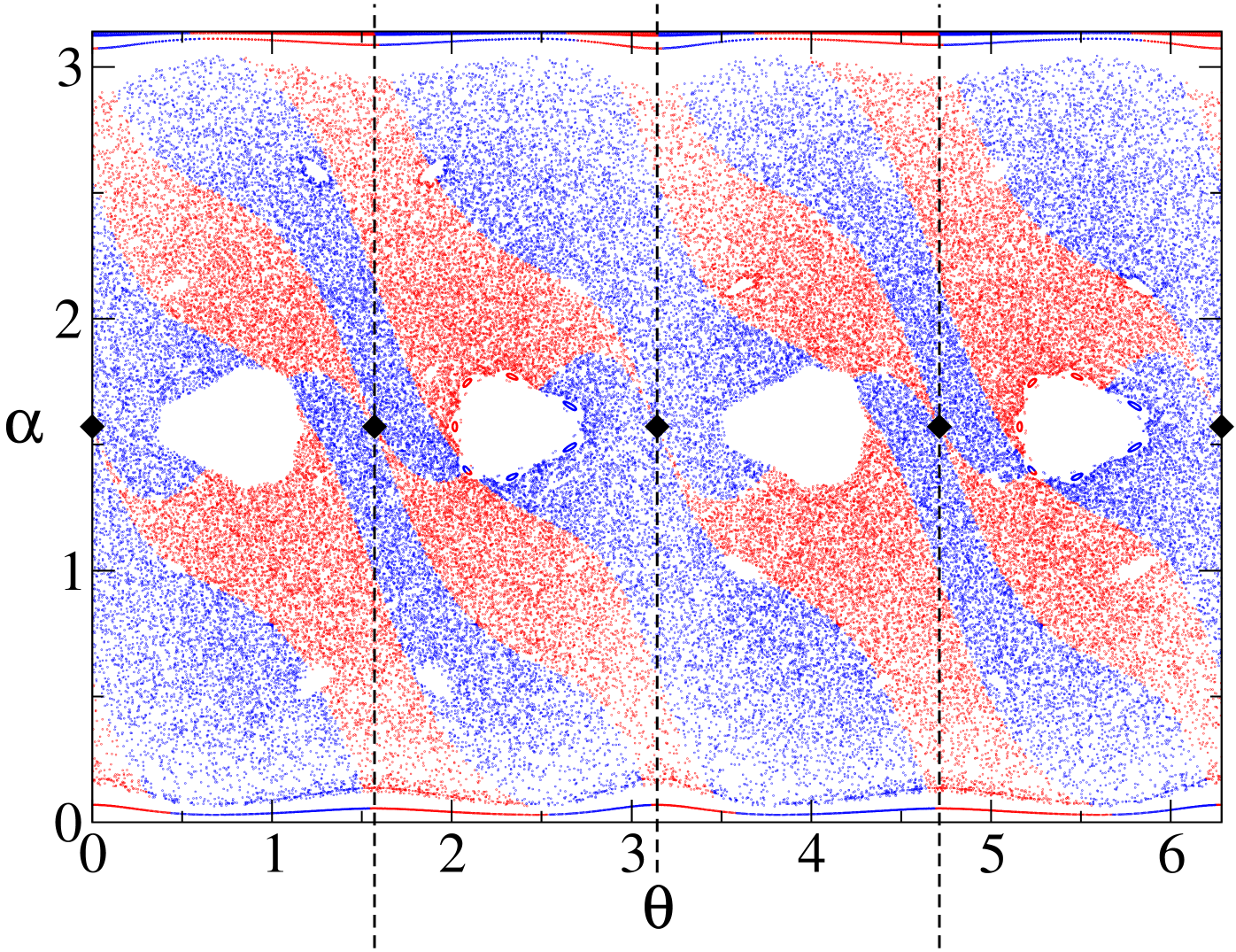}}    

    \raggedright (c)\vspace{-15pt}
    \centerline{\includegraphics[width=0.8\linewidth]{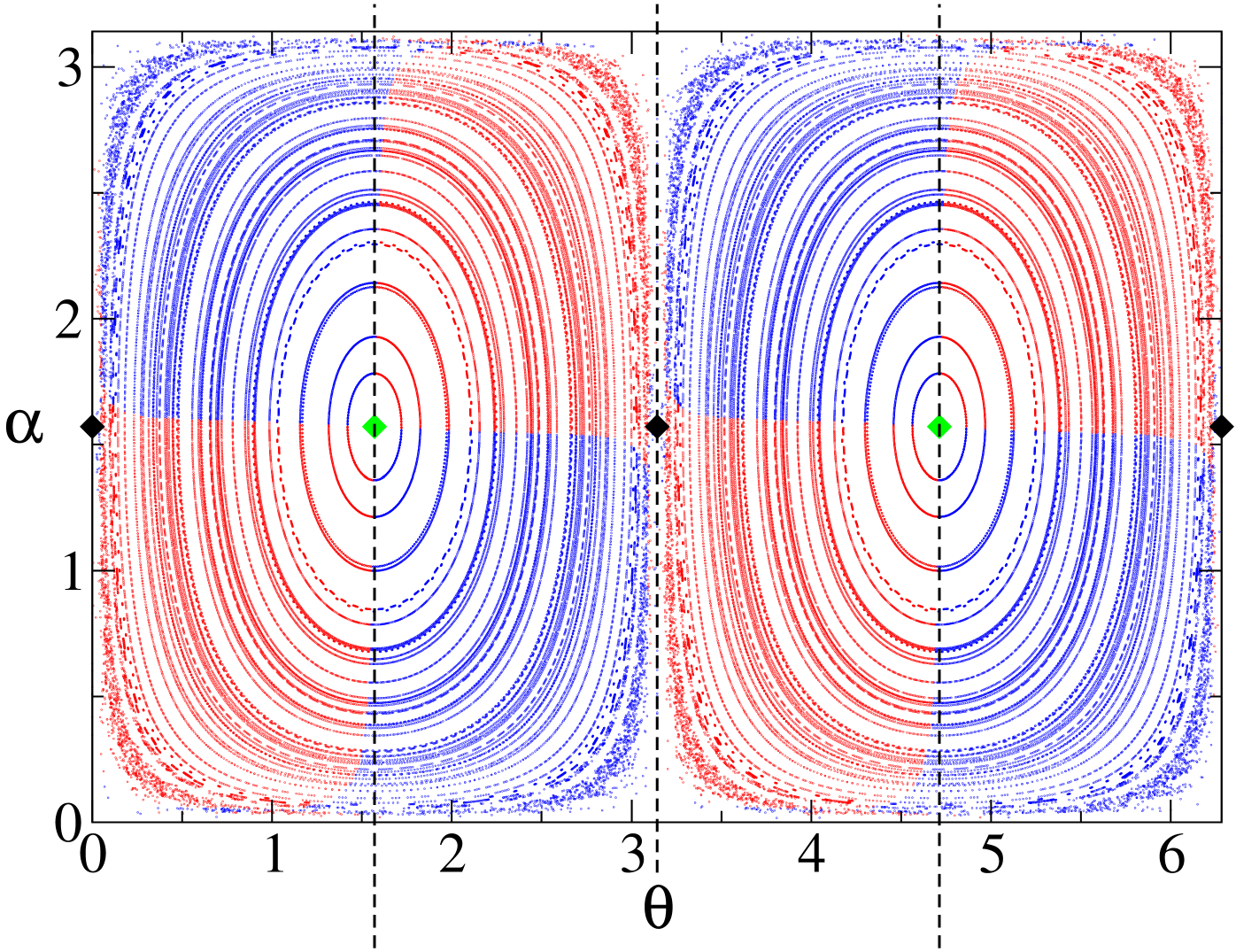}}  
\caption{Plot of the phase space for different $\varepsilon$ a)$\varepsilon=10$,b)$\varepsilon=0.5$ and c)$\varepsilon=0.01$. The figures were colored according to the noncanonical determinant $\det J$ (blue for $\det J<1$ and red for $\det J>1$). The markers indicate selected period-two points, which are fixed points of $T^2$ (green for elliptic and black for hyperbolic).}
\label{fig004}
\end{figure}

Since Eqs.~(9) and (10) are not generally equal to unity, the coordinate plane contains regions with $\det J>1$ and $\det J<1$. Displaying these values provides a diagnostic of the representation of the billiard map in $(\theta,\alpha)$ and helps reveal its periodic-orbit structure \cite{jacobian}. These colors must be interpreted in noncanonical coordinates; Liouville measure is preserved in both regions. Figure~\ref{fig004} illustrates this coordinate deformation for three values of $\varepsilon$.
The implicit collision map can make the direct numerical search for periodic orbits delicate. In the present problem, the Jacobian colors are a convenient way to visualize certain phase-space patterns. For an $N$-periodic trajectory, $T^N(\theta^*,\alpha^*)=(\theta^*,\alpha^*)$, the product of the $N$ Jacobian determinants equals one because the invariant-measure density returns to its initial value. Individual factors need not equal one. The marked axial and diagonal trajectories in the following figures are period-two orbits, hence fixed points of $T^2$, rather than interior fixed points of the single-collision map.

\begin{figure}[h!]
\centering
    \raggedright (a)
    \centerline{\includegraphics[width=0.9\linewidth]{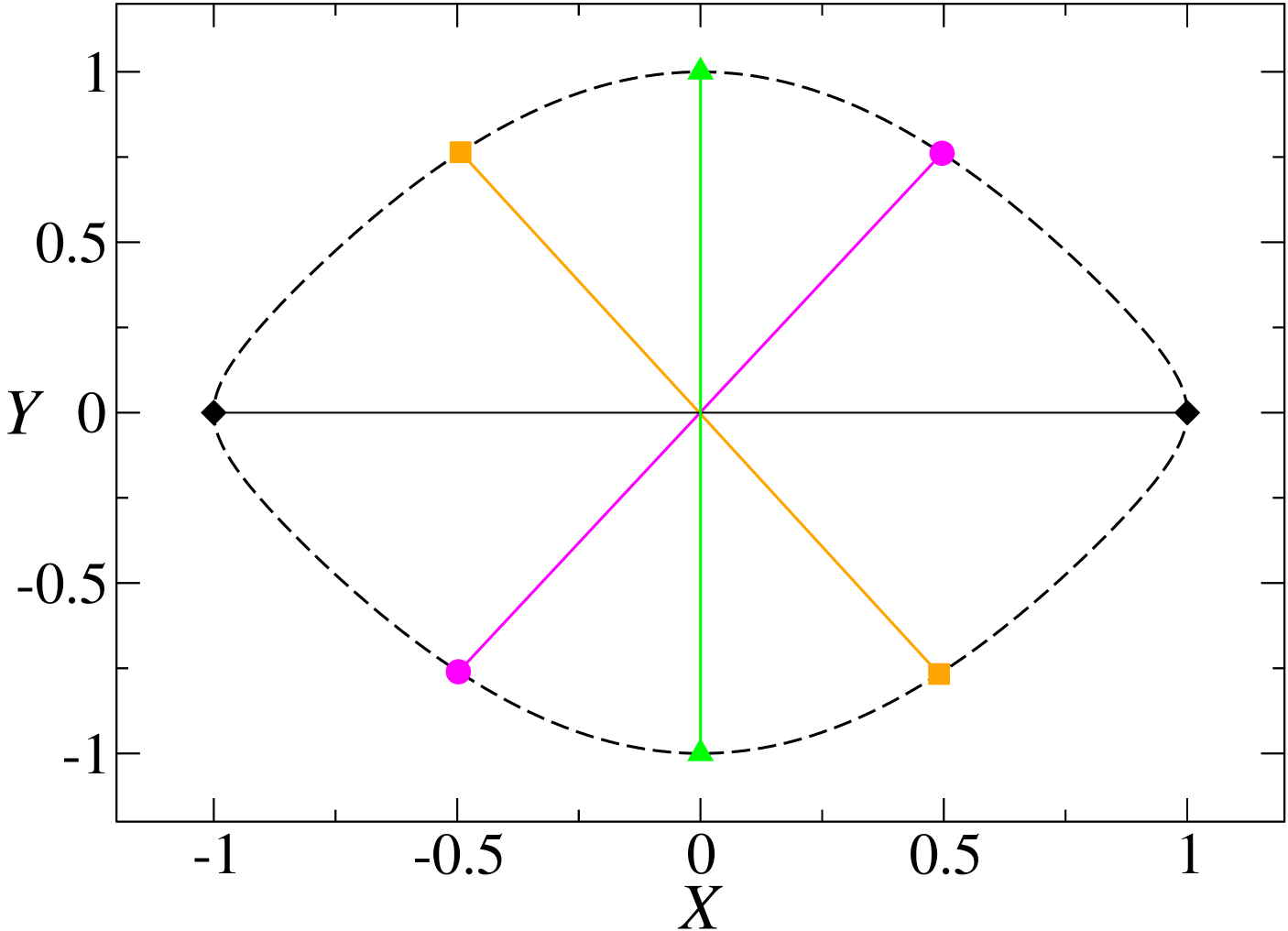}}
 
  \raggedright (b)
    \centerline{\includegraphics[width=0.9\linewidth]{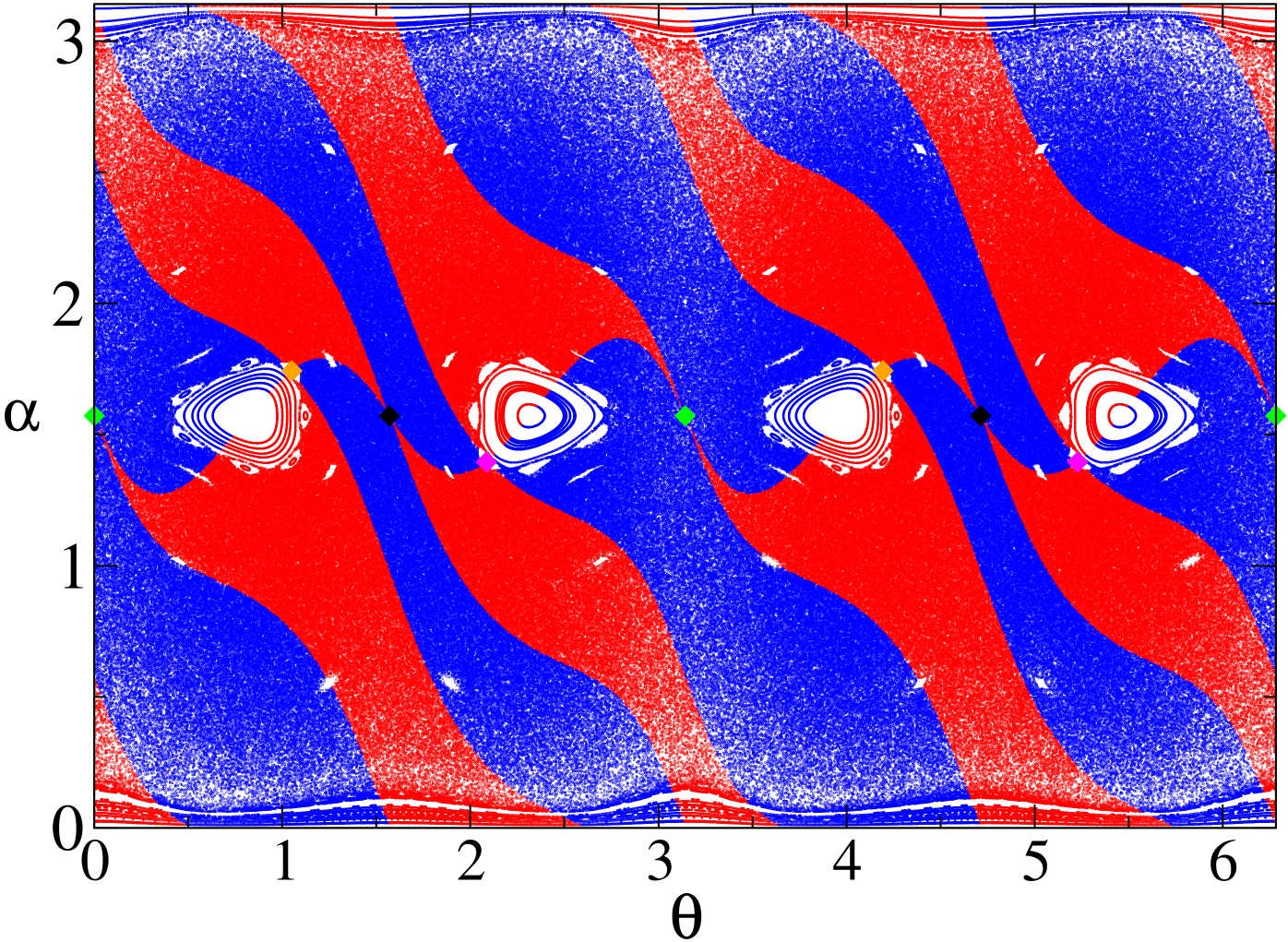}}    
\caption{(a) Period-two trajectories at $\varepsilon=0.5$: horizontal (black), vertical (cyan), and the analytically determined diagonal diameters (orange and magenta). (b) Noncanonical Jacobian colors (blue for $\det J<1$ and red for $\det J>1$), with markers locating the corresponding periodic points.}
\label{fig005}
\end{figure}

Figure~\ref{fig005} shows the axial period-two trajectories and two additional diagonal period-two orbits at $\varepsilon=0.5$, indicated in orange and magenta. Their existence can be determined analytically from the central inversion symmetry $\mathbf r(\theta+\pi)=-\mathbf r(\theta)$, with $\mathbf r=(X,Y)$. A trajectory connecting antipodal boundary points returns along the same diameter after reflection precisely when this diameter is perpendicular to the tangent. Hence $\mathbf r(\theta)\cdot\mathbf r'(\theta)=0$, giving
\begin{equation}
\sin\theta\cos\theta\left[-1+\varepsilon^2(1+\sin^2\theta)(1+3\sin^2\theta)\right]=0.
\label{eq_diagonal_condition}
\end{equation}
The factors $\sin\theta\cos\theta=0$ give the axial diameters. For nonaxial solutions, $u=\sin^2\theta$ satisfies
\begin{equation}
u=\frac{\sqrt{1+3/\varepsilon^2}-2}{3},\qquad 0<u<1.
\label{eq_diagonal_solution}
\end{equation}
It follows that two distinct diagonal diameters exist exactly when
\begin{equation}
\frac{1}{2\sqrt{2}}<\varepsilon<1.
\label{eq_diagonal_window}
\end{equation}
At the endpoints of this interval, the diagonal solutions merge with the corresponding axial diameters. Their linear stability requires a separate examination of the two-collision monodromy matrix.

The colored representation is a useful visual diagnostic, but orbital stability is determined by the trace of the two-step derivative $DT^2=J_2J_1$, not the value of a single-collision determinant. For the symmetric axial diameter orbits, the local linearizations at the two impacts have the same form in the adopted collision chart, allowing direct analytical evaluation.

In order to study the pair of fixed points located at  $\alpha=\pi/2$ and $\theta=\pi/2,3\pi/2$ we evaluate the behavior of the boundary in such points as $X(\pi/2)=0$, $Y(\pi/2)=2\varepsilon$, $X(3\pi/2)=0$ and $Y(3\pi/2)=-2\varepsilon$. The associated displacements are then $\Delta X=0$ and $\Delta Y=-4\varepsilon$. After calculating the derivatives $\frac{\partial \phi}{\partial \theta}$ and the individual terms of Eq.(8), we are able to obtain
\begin{equation}
    \text{Tr}(J) = j_{11} + j_{22} = 2(1 - 16\varepsilon^2)
\end{equation}
and for the period two 
\begin{equation}
    \text{Tr}(DT^2) = [\text{Tr}(J)]^2 - 2\det(J) = 4(1 - 16\varepsilon^2)^2 - 2
\end{equation}
since $\det J=1$ at this point. Now we can access the stability of these conditions as $-2 < \text{Tr}(DT^2) < 2$. The limit $Tr=2$ leads to the positive solution
\begin{equation}
    \varepsilon=\frac{1}{2\sqrt{2}}\approx 0.353553
\end{equation}
For $0<\varepsilon<1/(2\sqrt2)$, the vertical orbit is linearly elliptic away from the isolated marginal parameter $\varepsilon=1/4$, where $\text{Tr}(DT^2)=-2$. For $\varepsilon>1/(2\sqrt2)$, its trace exceeds two and the orbit is hyperbolic.
The same procedure for the fixed point in $\alpha=\pi/2$ and $\theta=0,\pi$ leads to 
\begin{equation}
    \text{Tr}(DT^2) = [\text{Tr}(J)]^2 - 2\det(J) = 4\left(\frac{2}{\varepsilon^2} - 1\right)^2 - 2
\end{equation}
which is linearly elliptic for $\varepsilon>1$, apart from the marginal parameter $\varepsilon=\sqrt2$ where $\text{Tr}(DT^2)=-2$, and hyperbolic for $0<\varepsilon<1$. The stability changes of the two axial diameters occur at $1/(2\sqrt2)$ and $1$, precisely the endpoints of the existence interval for the diagonal diameters in Eq.~(\ref{eq_diagonal_window}). The two thresholds differ because deformation changes the local curvature and the collision-to-collision distance along each diameter. The green and black markers in Fig.~\ref{fig004} identify the elliptic and hyperbolic axial periodic points, respectively.

\subsection{Lyapunov Exponents}

To investigate the presence of a chaotic component in parameter regimes where the phase-space structure is not immediately evident visually, such as near the singular geometric limit in Fig.~\ref{fig003}(a), we evaluate the maximal Lyapunov exponent associated with the mapping. This exponent also allows us to quantify how orbital instability varies with $\varepsilon$.

The Lyapunov exponents quantify the asymptotic exponential rates of growth or decay of infinitesimal perturbations along a trajectory in phase space. For a discrete-time two-dimensional mapping $(\theta_{n+1}, \alpha_{n+1}) = T(\theta_n, \alpha_n)$, the spectrum of Lyapunov exponents $\lambda_j$ ($j=1,2$) is formally defined as \cite{add1}
\begin{equation}
    \lambda_j = \lim_{n \to \infty} \frac{1}{n} \ln \Lambda_j(n),
    \label{eq_lyap_def}
\end{equation}
where $\Lambda_j(n)$ denote the singular values, ordered from largest to smallest, of the tangent propagator $M_n=J_n\cdots J_2J_1$, and $J_i$ is the local Jacobian evaluated along the trajectory. The exponents are measured per collision. This singular-value formulation is appropriate even when the finite-time propagator is nonnormal.

Direct numerical evaluation of Eq.~(\ref{eq_lyap_def}) by matrix multiplication presents severe computational limitations as $n$ increases. Exponential stretching and contraction make $M_n$ increasingly ill-conditioned, potentially causing numerical overflow and loss of the contracting direction. To address these limitations, we apply a QR decomposition at each iteration:
$J_i Q_{i-1} = Q_i T_i$,
where $Q_i$ is an orthogonal matrix and $T_i$ is an upper triangular matrix with positive diagonal entries $T_{jj}^i$. Starting from a generic orthonormal basis $Q_0$, the Lyapunov exponents are evaluated from the accumulated logarithms of these diagonal entries:
\begin{equation}
    \lambda_j = \lim_{n \to \infty} \frac{1}{n} \sum_{i=1}^n \ln \left| T_{jj}^i \right|, \quad j=1,2.
    \label{eq_lyap_qr}
\end{equation}
In canonical billiard coordinates, $\det M_n=1$. In the noncanonical variables used here, the Jacobian factors telescope according to Eq.~(10), giving
$\det M_n=s'(\theta_0)\sin\alpha_0/[s'(\theta_n)\sin\alpha_n]$,
rather than exactly unity at every finite $n$. Provided the logarithm of this endpoint factor grows sublinearly with $n$, its contribution to the asymptotic average vanishes, yielding $\lambda_1+\lambda_2=0$. The maximal exponent $\lambda\equiv\lambda_1$ characterizes exponential orbital instability when positive, while regular quasiperiodic trajectories have $\lambda=0$ in the asymptotic limit. A sum approaching zero provides a useful numerical consistency check under the stated assumption.

\begin{figure*}[htbp]
    \centering
    \begin{minipage}[t]{0.32\linewidth}
        \raggedright (a)\\
        \centering
        \includegraphics[width=\textwidth]{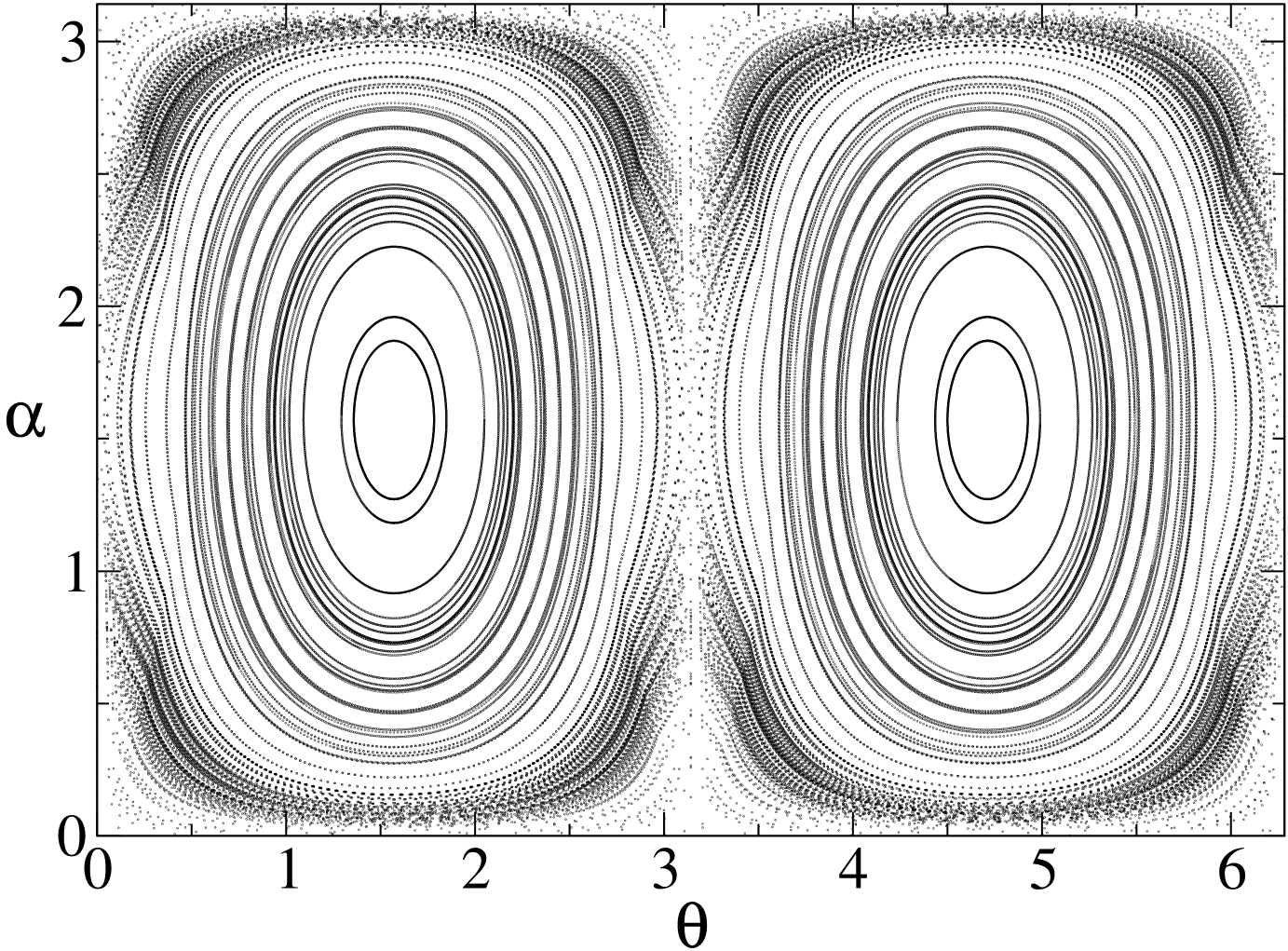}
    \end{minipage}%
    \hfill
    \begin{minipage}[t]{0.32\linewidth}
        \raggedright (b) \\
        \centering 
        \includegraphics[width=\textwidth]{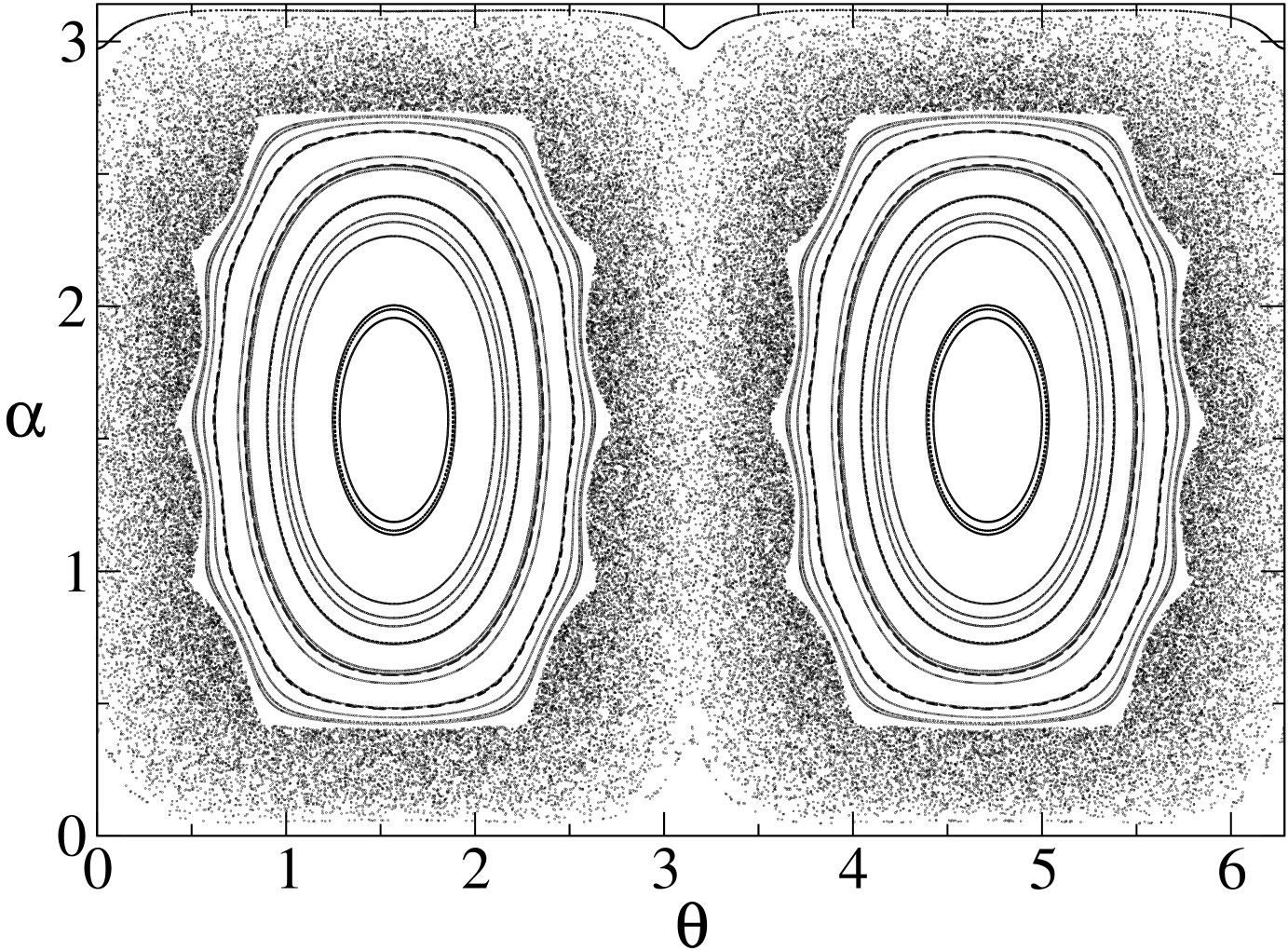}
    \end{minipage}%
    \hfill
    \begin{minipage}[t]{0.32\linewidth}
        \raggedright (c) \\
        \centering 
        \includegraphics[width=\textwidth]{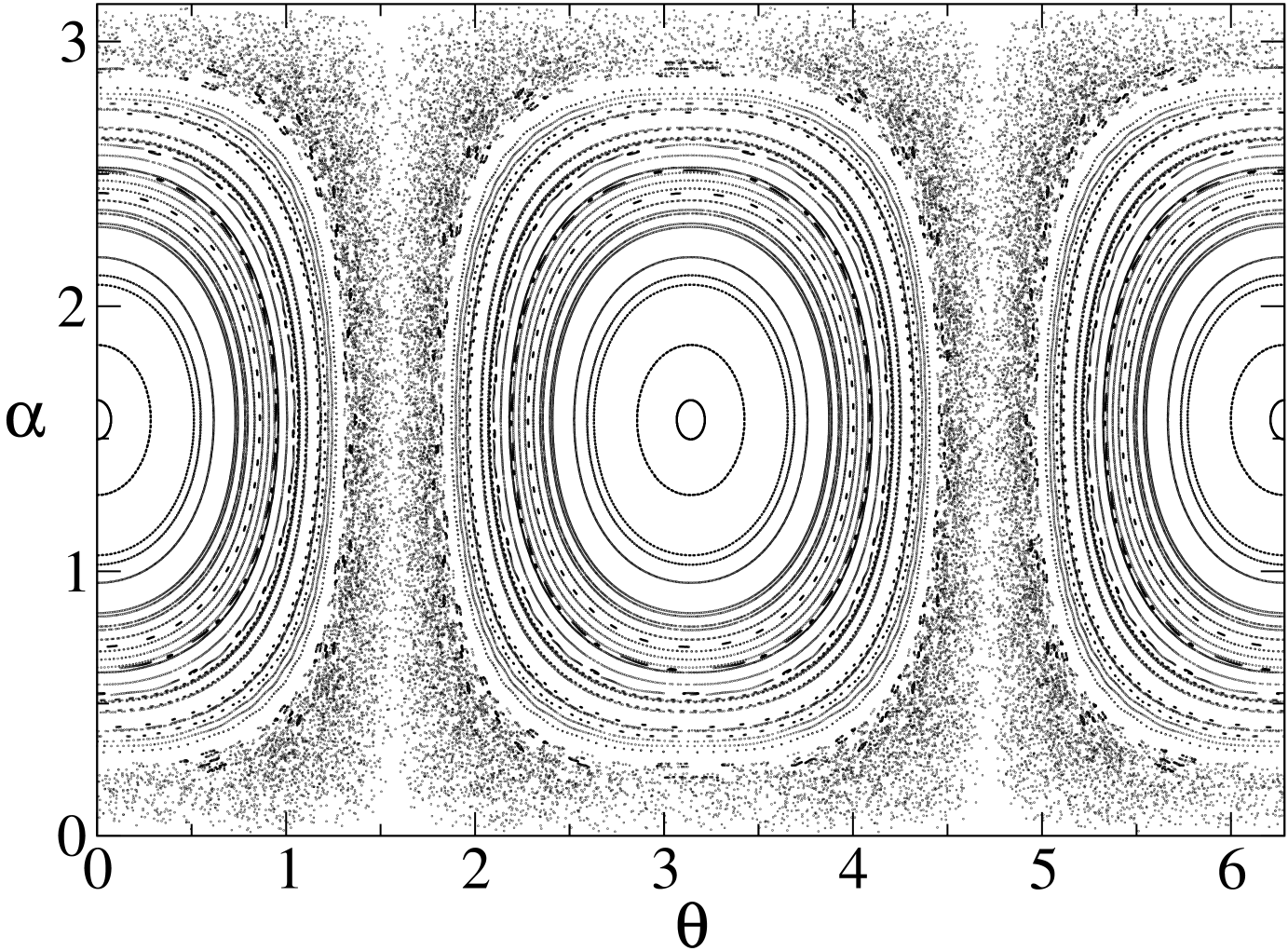}
    \end{minipage}%

    \vspace{0.3cm} %

    \begin{minipage}[t]{0.32\linewidth}
        \raggedright (d) \\
        \centering 
        \includegraphics[width=\textwidth]{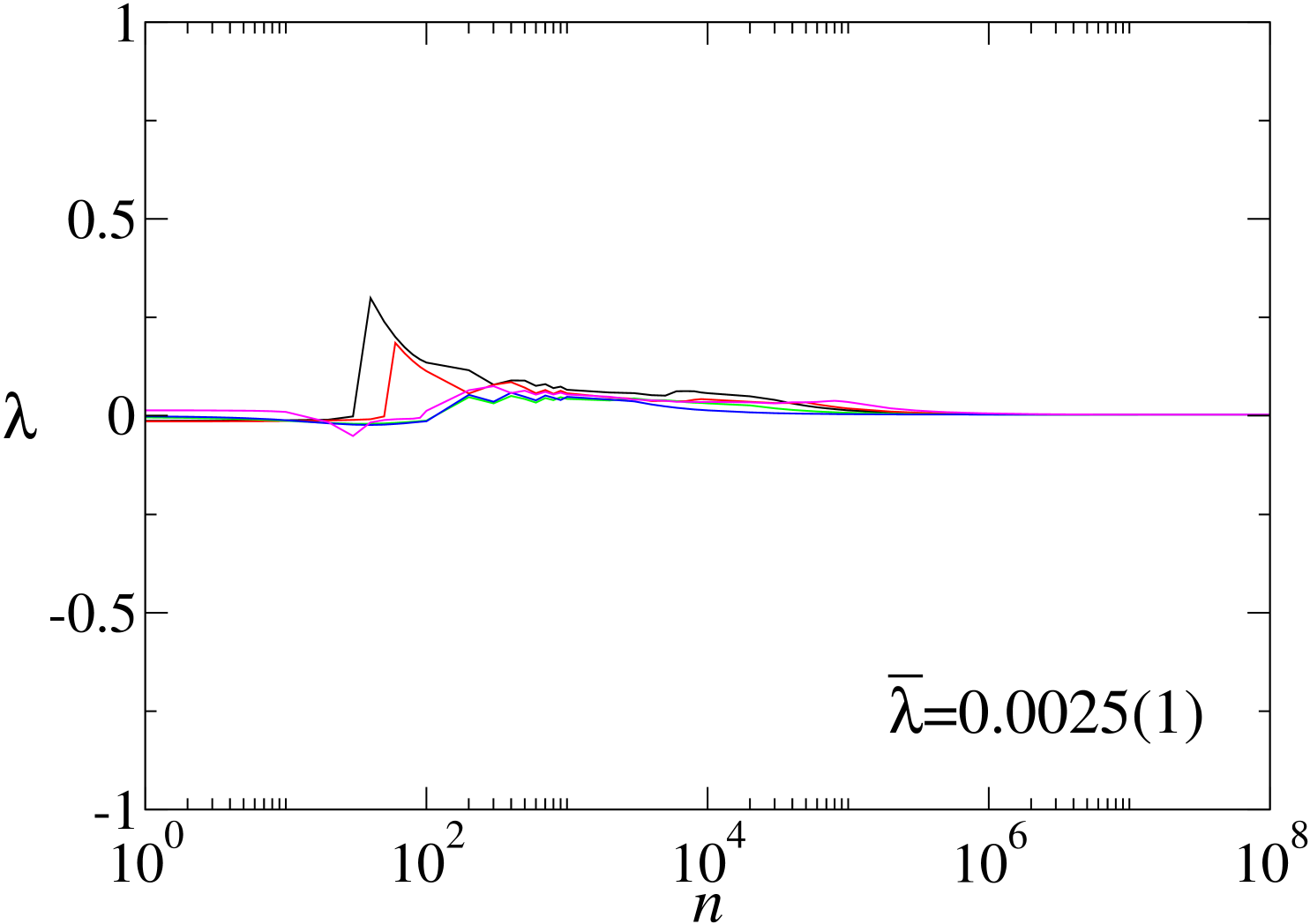}
    \end{minipage}%
    \hfill
    \begin{minipage}[t]{0.32\linewidth}
       \raggedright (e) \\
        \centering 
        \includegraphics[width=\textwidth]{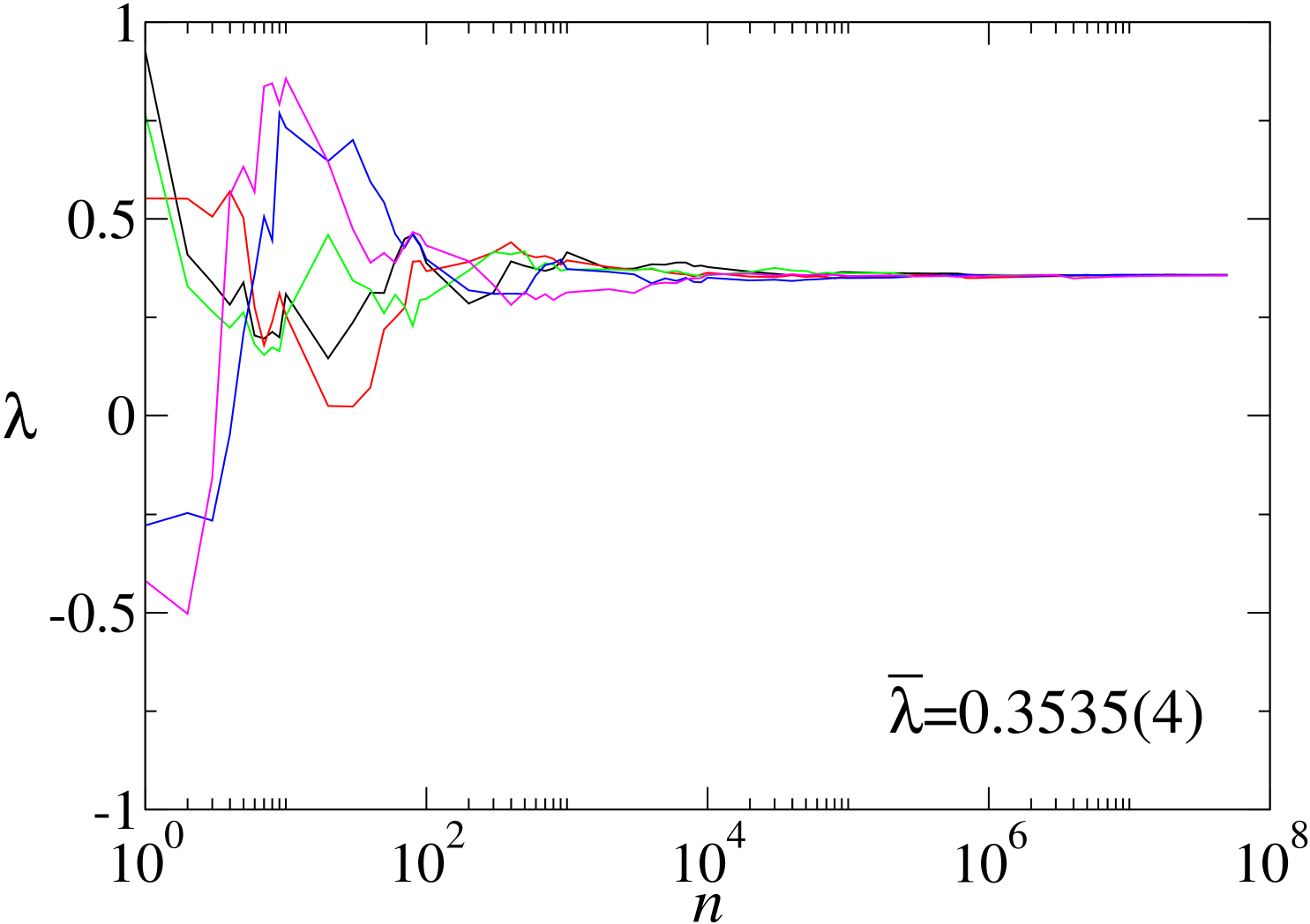}
    \end{minipage}%
    \hfill
    \begin{minipage}[t]{0.32\linewidth}
        \raggedright (f) \\
        \centering 
        \includegraphics[width=\textwidth]{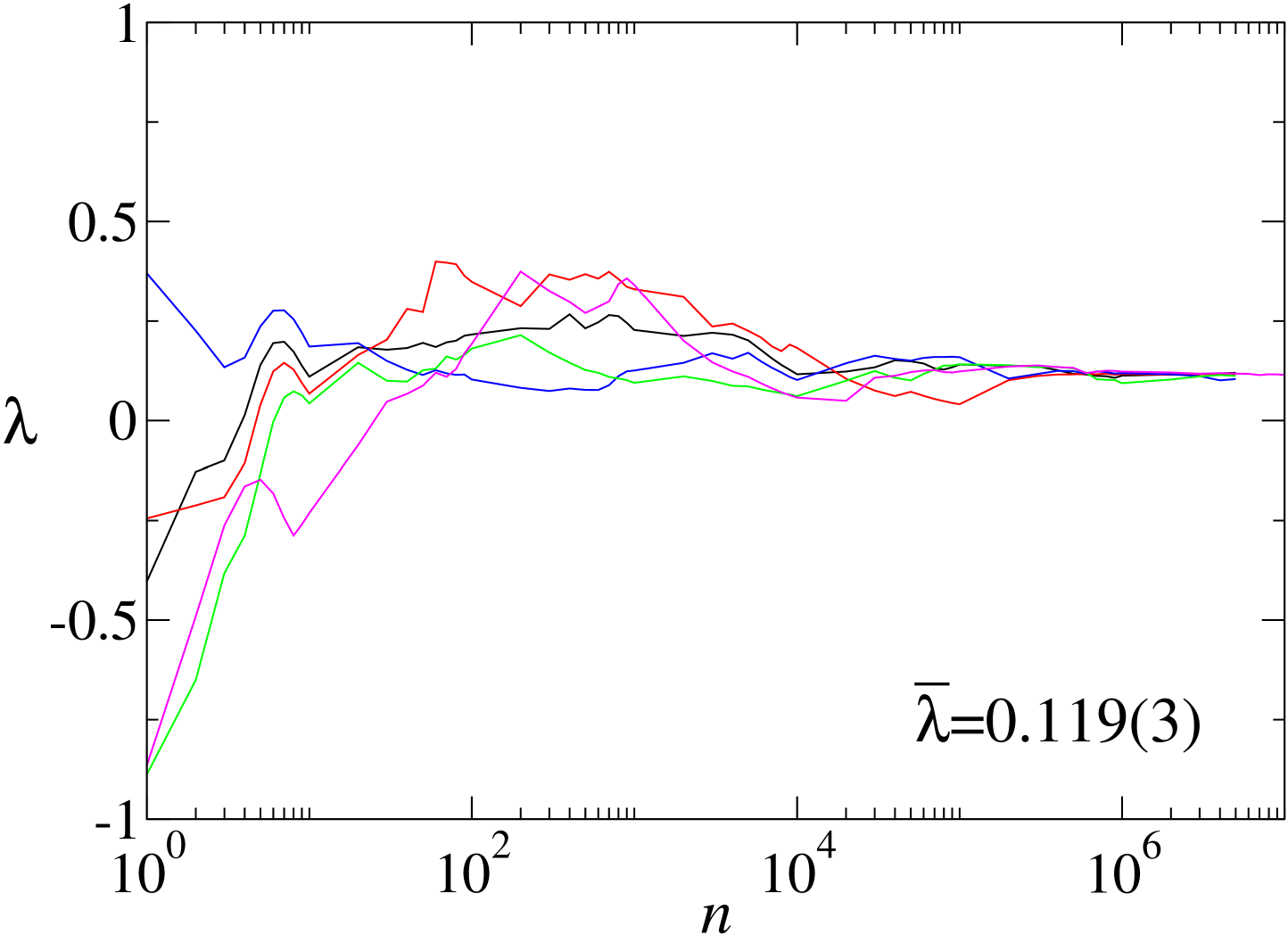}
    \end{minipage}%

    \vspace{0.3cm}

    \caption{Finite-time estimates of the maximal Lyapunov exponent $\lambda$ versus collision number $n$ for five representative chaotic trajectories at each value of $\varepsilon$: $\varepsilon=0.001$ for the first column in panels (a) and (d), $\varepsilon=0.1$ for the second in panels (b) and (e) and $\varepsilon=10$ in the last column, for panels (c) and (f). The average plateau estimate $\overline{\lambda}$ is displayed inside each panel.}
    \label{figlyap}
\end{figure*}

Figure~\ref{figlyap} presents finite-time estimates of the maximal Lyapunov exponent $\lambda$ as a function of collision number $n$ for the values of $\varepsilon$ examined above. For each case, five initial conditions were selected on the line $\alpha=3.0$, with $\theta\in[0,2\pi]$, and trajectories were iterated up to $n=10^8$ collisions. The resulting curves approach positive plateaus, providing evidence of chaotic components at the tested parameter values. Several curves exhibit transient downward excursions in the cumulative estimate $\lambda(n)$, consistent with dynamical stickiness near regular islands \cite{Bunimovich_2008}. During a long near-regular episode, the accumulated average in Eq.~(\ref{eq_lyap_qr}) can decrease; a subsequent chaotic episode can raise it again. Assessing convergence to an asymptotic exponent and distinguishing sticky transients from numerical nonconvergence require trajectory-by-trajectory checks and additional independent initial conditions.

\begin{figure}[h!]
    \centering
    \includegraphics[width=\linewidth]{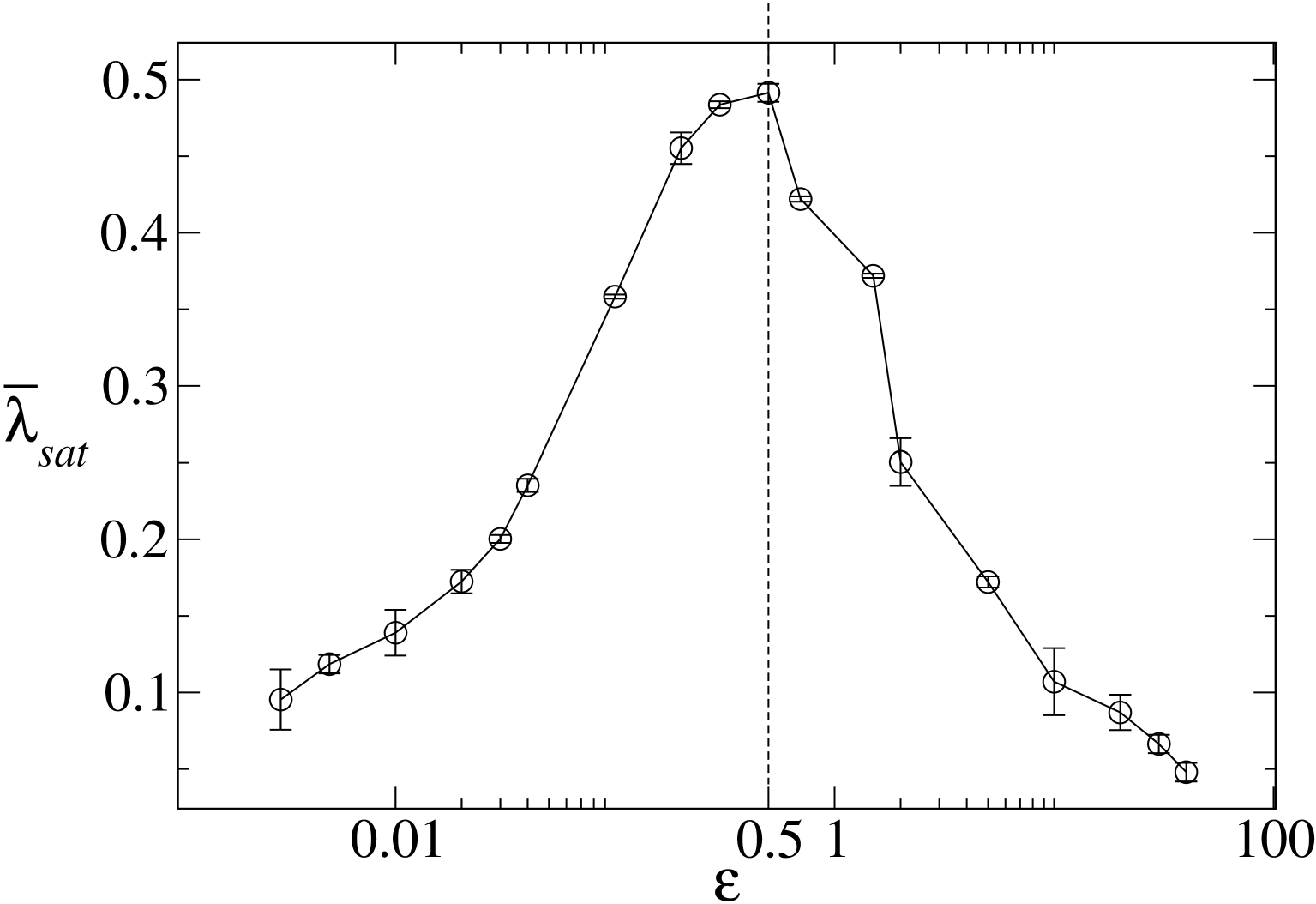}
    \caption{Average finite-time plateau value of the maximal Lyapunov exponent $\bar{\lambda}_{\text{sat}}$ as a function of $\varepsilon$ (logarithmic scale). Error bars denote the standard deviation across five initial conditions selected in the sampled chaotic region. The dashed line marks the largest value within the parameter scan, near $\varepsilon=0.5$.}
    \label{figlyap_sat}
\end{figure}

Analyzing the average plateau value of the maximal Lyapunov exponent, $\bar{\lambda}_{\text{sat}}$, as a function of the control parameter quantifies the nonmonotonic variation of sampled orbital instability between the two singular geometric limits. Figure~\ref{figlyap_sat} displays $\bar{\lambda}_{\text{sat}}$ across five orders of magnitude of $\varepsilon$, from $\varepsilon=10^{-3}$ to $\varepsilon=10^2$, obtained by averaging over five independent chaotic trajectories.

The scan exhibits a prominent peak near $\varepsilon=0.5$ (vertical dashed line), where the reported plateau value is $\bar{\lambda}_{\text{sat}}\approx0.491(5)$. Within the explored parameter ranges, this estimate decreases toward both singular geometric limits. The flattening boundary at small $\varepsilon$ and elongated boundary at large $\varepsilon$ substantially change the collision geometry, but the finite scan alone does not establish an integrable limiting billiard or a unique global maximum of instability.

This nonmonotonic dependence indicates that orbital instability along the sampled chaotic trajectories is strongest at an intermediate geometry, near $\varepsilon=0.5$, where the horizontal and vertical semiaxes have equal lengths. The Lyapunov exponent characterizes instability along selected trajectories rather than the phase-space measure of the chaotic component, and a positive exponent alone does not establish rapid global mixing. We therefore discuss the extent of the chaotic envelope separately in Sec.~\ref{xsec4}.

\subsection{Invariant curves}

As the control parameter $\varepsilon$ is varied, the overall topological structure of the phase space undergoes significant transformations. While increasing $\varepsilon$ enhances the non-linearity of the system and promotes the growth of chaotic seas around resonant island chains, it does not lead to global, unbounded chaos. Instead, robust invariant curves persist near the grazing incidence limits ($\alpha \to 0$ and $\alpha \to \pi$) across all finite values of $\varepsilon > 0$. To understand the physical and geometric origins of these persistent regular structures, we evaluate the global differential geometry of the boundary and apply Lazutkin's theorem \cite{lazutkin} within the framework of KAM theory.

The local curvature $\kappa(\theta)$ of a parametrically defined boundary $[X(\theta), Y(\theta)]$ is given by
\begin{equation}
    \kappa(\theta) = \frac{X'(\theta)Y''(\theta) - X''(\theta)Y'(\theta)}{\left[X'^2(\theta) + Y'^2(\theta)\right]^{3/2}}.
    \label{eq_curvature}
\end{equation}
Substituting the parametric derivatives of Eq.~(\ref{eq12}) into Eq.~(\ref{eq_curvature}), the numerator simplifies to $\varepsilon \left( 6\sin^4\theta - 3\sin^2\theta + 1 \right)$. Since the quadratic expression $6u^2 - 3u + 1 > 0$ holds for all $u = \sin^2\theta \in [0, 1]$, the numerator is positive-definite. Furthermore, the denominator remains bounded and strictly non-zero for any finite $\varepsilon > 0$, ensuring that
\begin{equation}
    \kappa(\theta) > 0 \quad \forall \theta \in [0, 2\pi], \; \forall \varepsilon > 0.
\end{equation}
This establishes that the boundary is $C^\infty$-smooth and strictly convex for every finite $\varepsilon>0$. Strict convexity alone does not exclude chaotic defocusing through successive reflections; the key consequence here is the persistence of sufficiently near-grazing caustics under the regularity conditions below.

The smoothness and strict convexity satisfy the hypotheses of Lazutkin's theorem \cite{lazutkin}, which ensures a positive-measure Cantor family of invariant caustics accumulating at the billiard boundary $\partial Q$. A caustic is a curve $\Gamma$ such that a ray tangent to it remains tangent after successive specular reflections. In the collision phase space, these caustics correspond to noncontractible invariant curves spanning $\theta\in[0,2\pi]$, associated with whispering-gallery trajectories at small incidence angles. They persist for each fixed finite $\varepsilon>0$, even though the theorem alone does not determine their total width as the geometry degenerates.

From a KAM-theoretic perspective, each surviving caustic is associated with an irrational rotation number $\omega$ satisfying the Diophantine condition:
\begin{equation}
    \left| \omega - \frac{p}{q} \right| \ge \frac{C}{q^\tau},
\end{equation}
where $C > 0$ and $\tau \ge 2$ are constants, and $p/q$ represents rational approximations. Because the set of Diophantine rotation numbers has positive Lebesgue measure and continuum cardinality, an infinite family of invariant caustics survives for every finite value of $\varepsilon > 0$ \cite{lazutkin}. Consequently, these invariant spanning curves act as topological barriers in the two-dimensional phase space, preventing chaotic trajectories in the interior from reaching the extreme grazing limits ($\alpha \to 0, \pi$) and ruling out global phase-space transport.

In the numerical portraits close to the singular limits, the rotational region near grazing incidence appears progressively narrow relative to the prominent libration domains. This visual observation should not be confused with a theorem about the asymptotic width of the surviving Lazutkin region: strict convexity ensures its existence for each finite $\varepsilon$, but does not supply an $\varepsilon$-dependent width estimate. The compressed appearance in Fig.~\ref{fig002} is consistent with the singular geometric limits described above.

This organization differs from the circle-to-oval deformation investigated in Ref.~\cite{separatriz}, where the perturbation starts from a regular, nondegenerate circular billiard. Here the reference geometries are singular line segments, and surviving boundary caustics coexist with a chaotic component throughout the finite-parameter family. This distinction motivates examining whether the asymptotic scaling of the chaotic envelope reflects the special geometry of the eye rather than the standard weak-perturbation mechanism of the oval billiard. Establishing a universality class across different models would require additional evidence beyond the single geometric family studied here.

\section{Scaling properties and critical exponents}
\label{xsec4}

The magnifications in Fig.~\ref{fig003} illustrate the chaotic component situated between libration islands and rotational invariant curves. For small $\varepsilon$, the vertical period-two orbit is elliptic and the horizontal one hyperbolic, according to Sec.~\ref{xsec3}; their stability roles reverse in the opposite asymptotic regime. The stochastic layer lies near the interfaces separating these qualitatively different motions.

As the geometry changes, regular islands and chaotic regions reorganize nonmonotonically. To quantify this change, we measure the angular extent of an envelope built from sampled chaotic trajectories. This is a geometric support diagnostic, not a diffusion coefficient or the canonical area of the chaotic component. We distinguish its convergence with increasing sampling from dynamical spreading of an individual trajectory in collision time.

\begin{figure}[h!]
\centering
    \raggedright (a)
    \centerline{\includegraphics[width=0.9\linewidth]{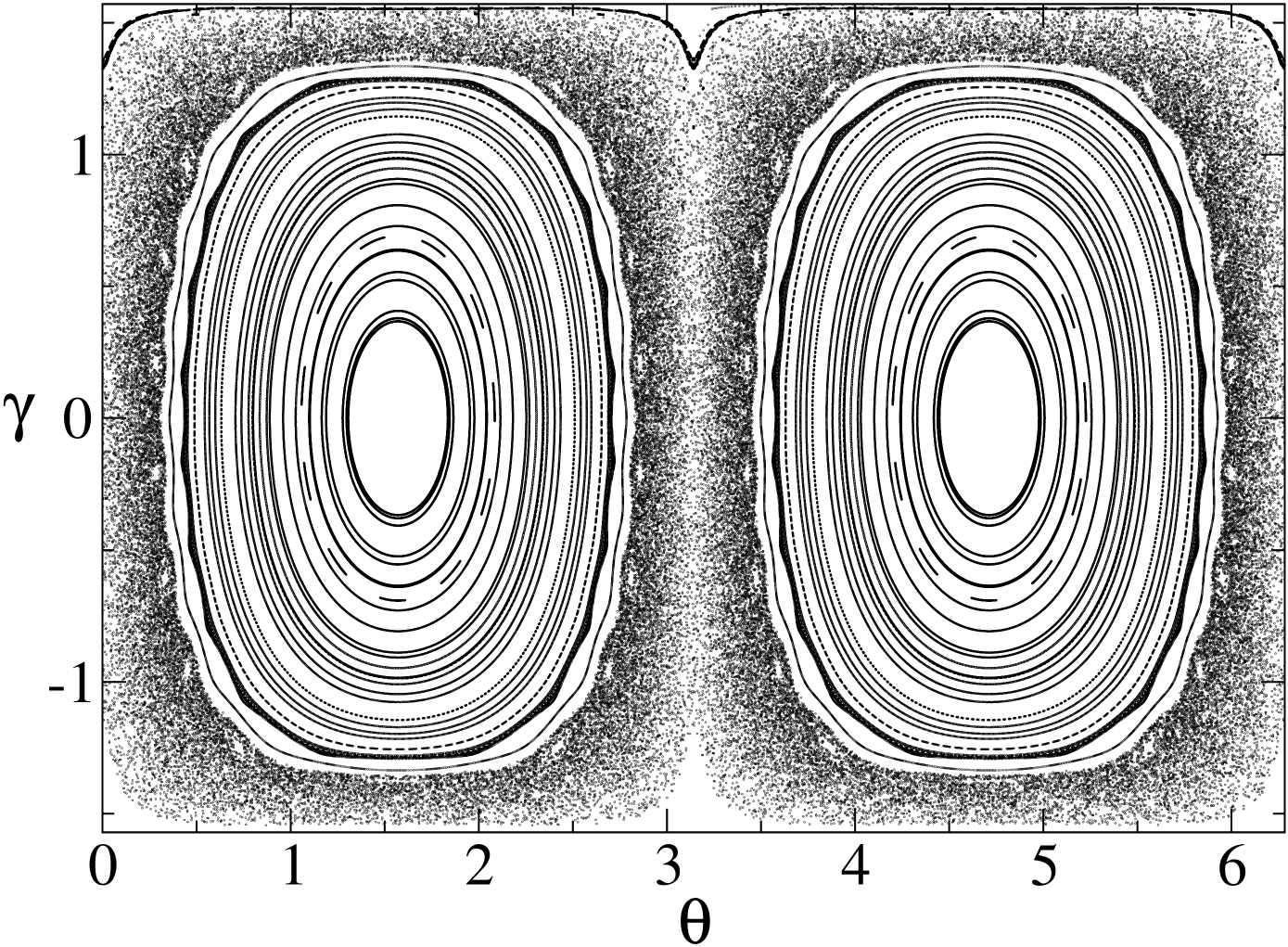}}
 
  \raggedright (b)
    \centerline{\includegraphics[width=0.9\linewidth]{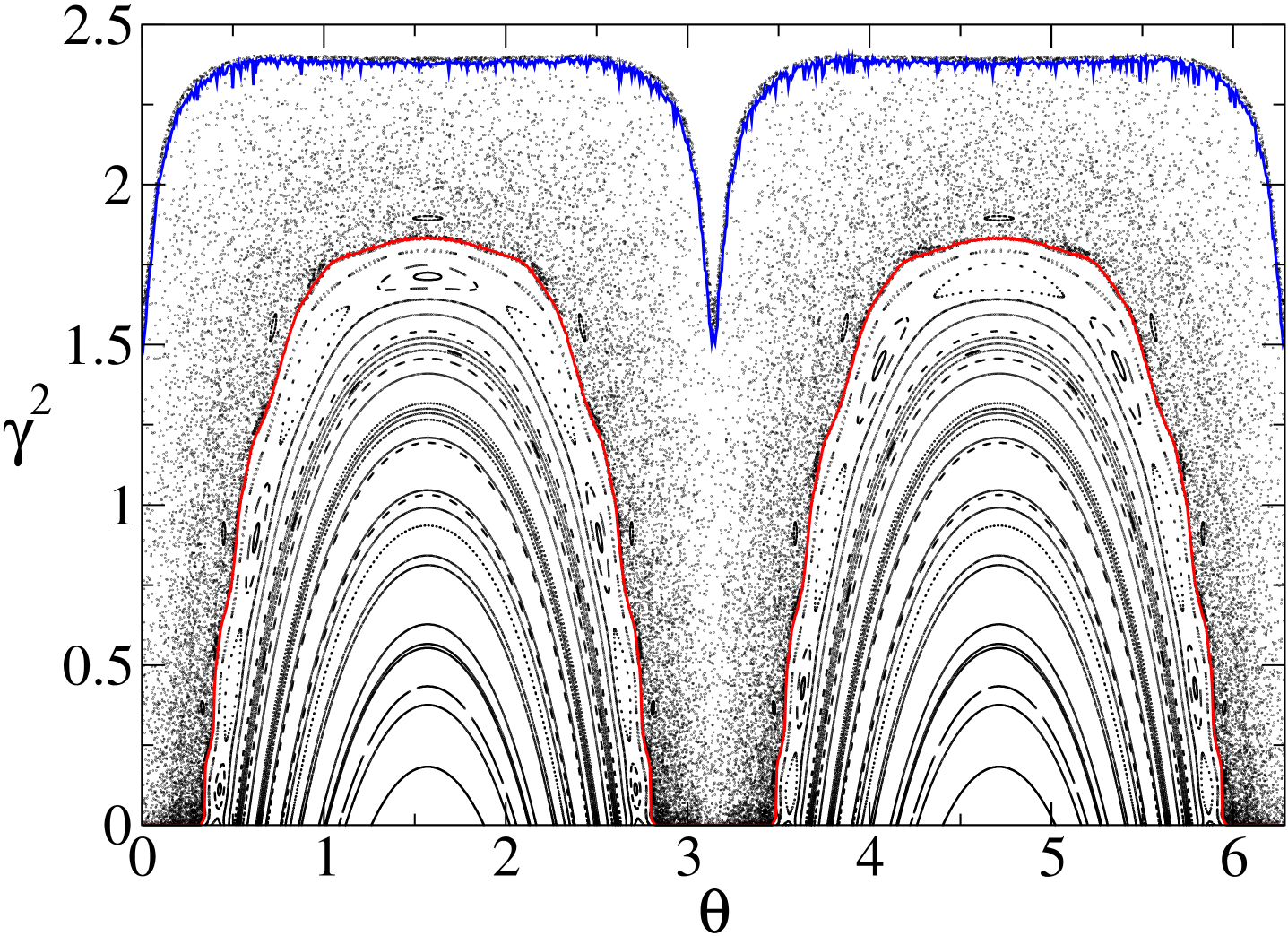}}  
\caption{(a) Collision phase space for $\varepsilon=0.05$ in the shifted coordinate $\gamma=\alpha-\pi/2$. (b) The same data folded through $\gamma\mapsto\gamma^2$. The blue and red curves record the sampled upper and lower extremes in angular bins after $n=10^8$ collisions for one chaotic trajectory. The envelope can include regular islands between its extrema and must not be identified with the actual chaotic area.}
\label{fig8}
\end{figure}

Fig.~\ref{fig8}(a) shows the phase-space portrait of
the eye billiard for  $\varepsilon=0.05$ after the
transformation $\gamma=\alpha-\pi/2$, which makes the symmetry of a well-developed chaotic layer explicit with respect to the origin $\gamma=0$.

The change $\gamma\mapsto\gamma^2$ folds the two signs of the shifted angle and highlights the extent of the sampled set; it does not establish statistical equivalence between them. Figure~\ref{fig8}(b) displays the resulting extrema. Their difference defines an \emph{envelope span}: points between the curves need not belong to the same chaotic component, and regular islands can occupy part of the enclosed region.

To track this sampled angular extent, we construct the average envelope width $\bar{W}$. The coordinate $\theta\in[0,2\pi]$ is partitioned into $M=1000$ bins of width $\Delta\theta=2\pi/M$. For an ensemble of trajectories selected from the chaotic component, collisions in bin $i$ update the extrema $\gamma^2_{\min,i}$ and $\gamma^2_{\max,i}$.

\begin{figure}[h!]
\centering
     \raggedright (a)
    \centerline{\includegraphics[width=0.9\linewidth]{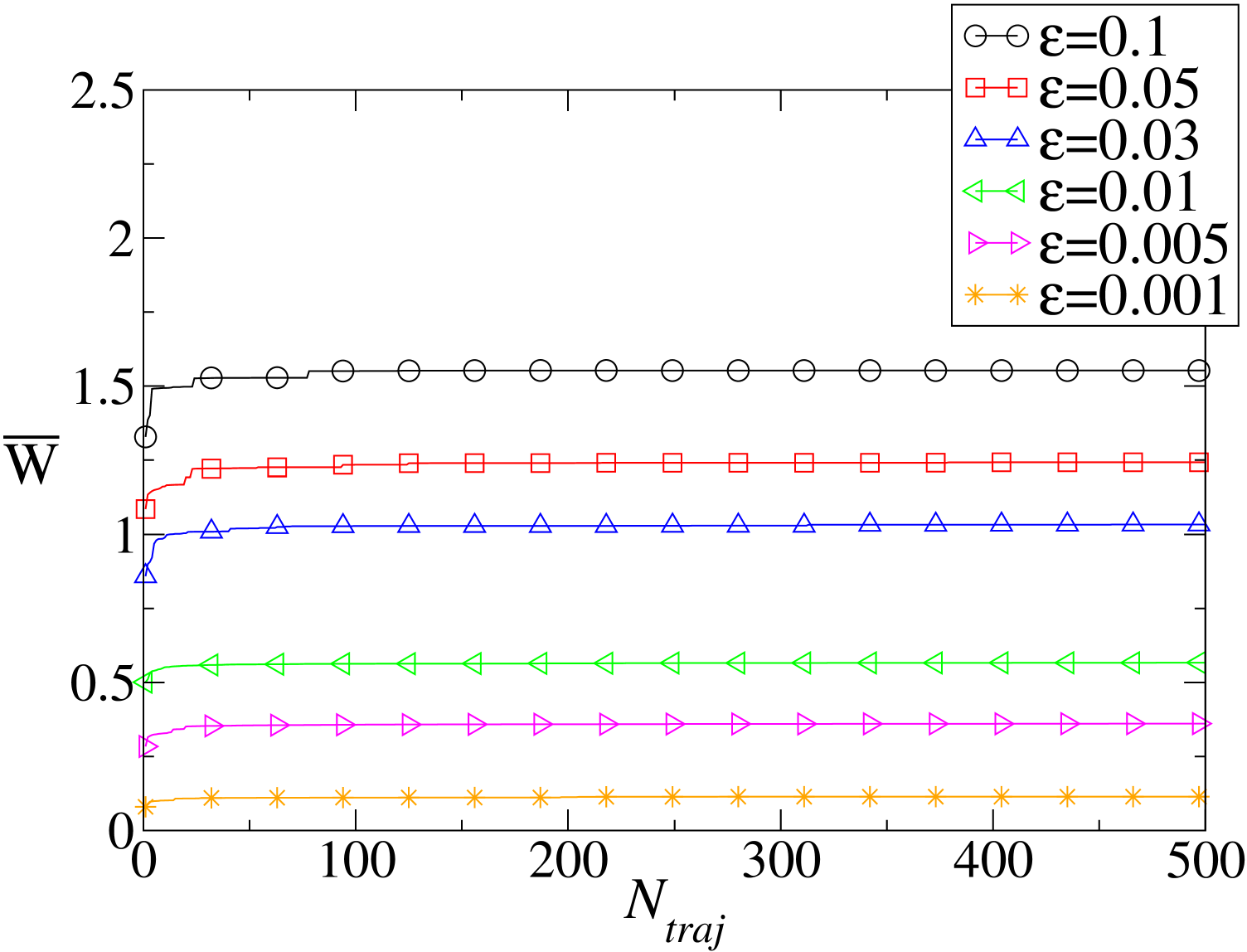}}
    
     \raggedright (b)
     \centerline{\includegraphics[width=0.9\linewidth]{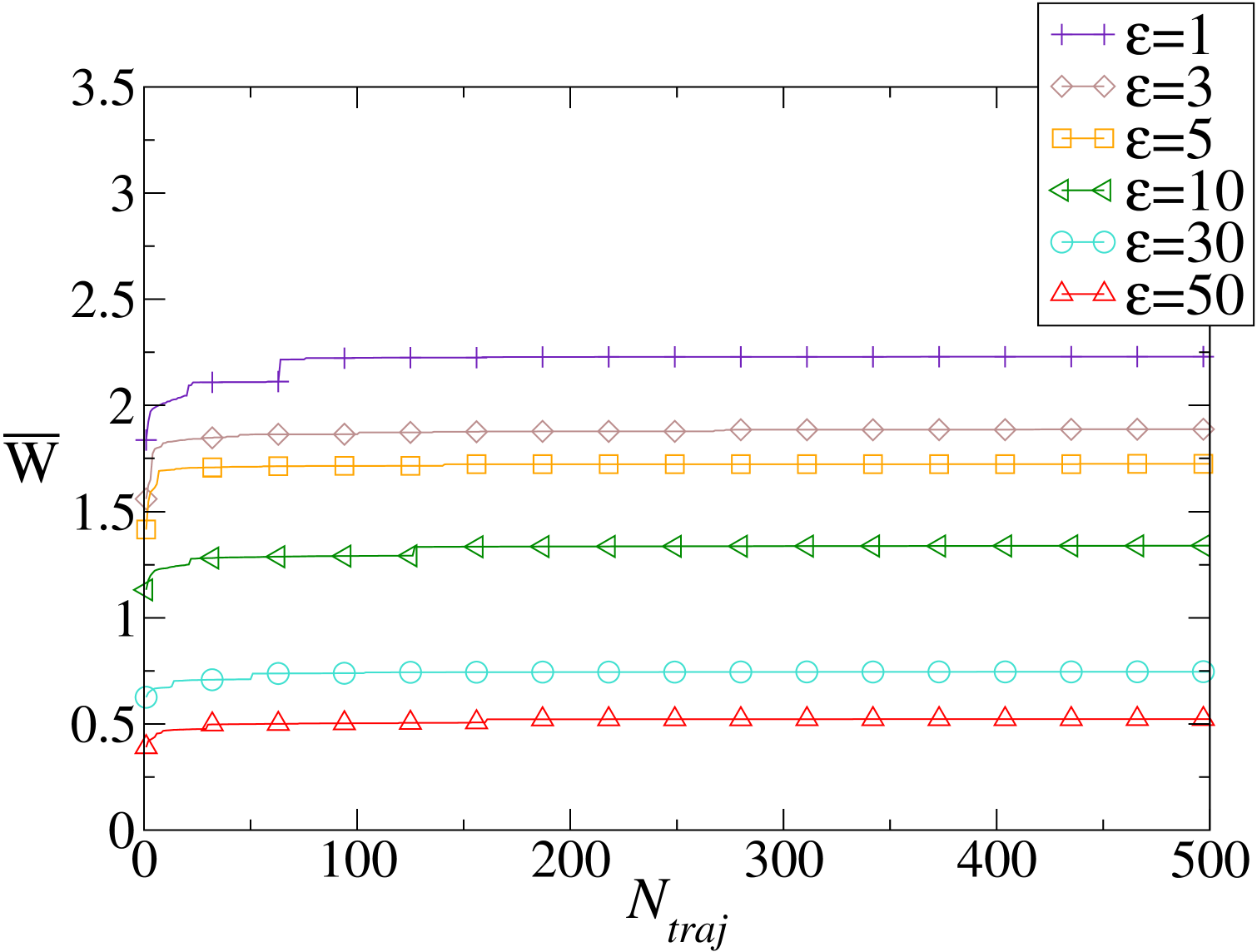}}    
\caption{Plot of $\overline{W}$ as a function of the number of trajectories $N_{traj}$ for different values of the control parameter $\varepsilon$, as labeled in the figure. Panel a) shows the decrease of the saturation value as $\varepsilon\rightarrow0$ while panel b) exhibits the same decrease as $\varepsilon \rightarrow \infty$.}
\label{fig9}
\end{figure}

The local thickness of the chaotic stripe within bin $i$ is defined as $W_i = \gamma^2_{\max, i} - \gamma^2_{\min, i}$. Averaging $W_i$ over all populated bins yields the global average envelope width:
\begin{equation}
    \bar{W} = \frac{1}{N_{valid}} \sum_{i=1}^{N_{valid}} \left( \gamma^2_{\max, i} - \gamma^2_{\min, i} \right),
    \label{eq_wbar}
\end{equation}
where $N_{\mathrm{valid}}\leq M$ counts populated bins. The normalization gives a conditional average over visited bins; before complete angular coverage, it is not necessarily an unbiased estimate of the full-angle envelope. Figure~\ref{fig9} plots $\bar{W}$ against the number $N_{\mathrm{traj}}$ of accumulated initial conditions, each iterated for $10^6$ collisions. The numerical saturation criterion used here is $N_{\mathrm{valid}}=M$ together with a relative change below $0.01\%$ over the last $100$ added trajectories. We denote the resulting plateau by $\bar{W}_{\mathrm{sat}}$. Because the observable is based on extrema, its limiting value and fitted exponent should additionally be checked against longer individual trajectories, independent ensembles, and changes in $M$.

\begin{figure}[h!]
\centering
     \raggedright (a)
    \centerline{\includegraphics[width=\linewidth]{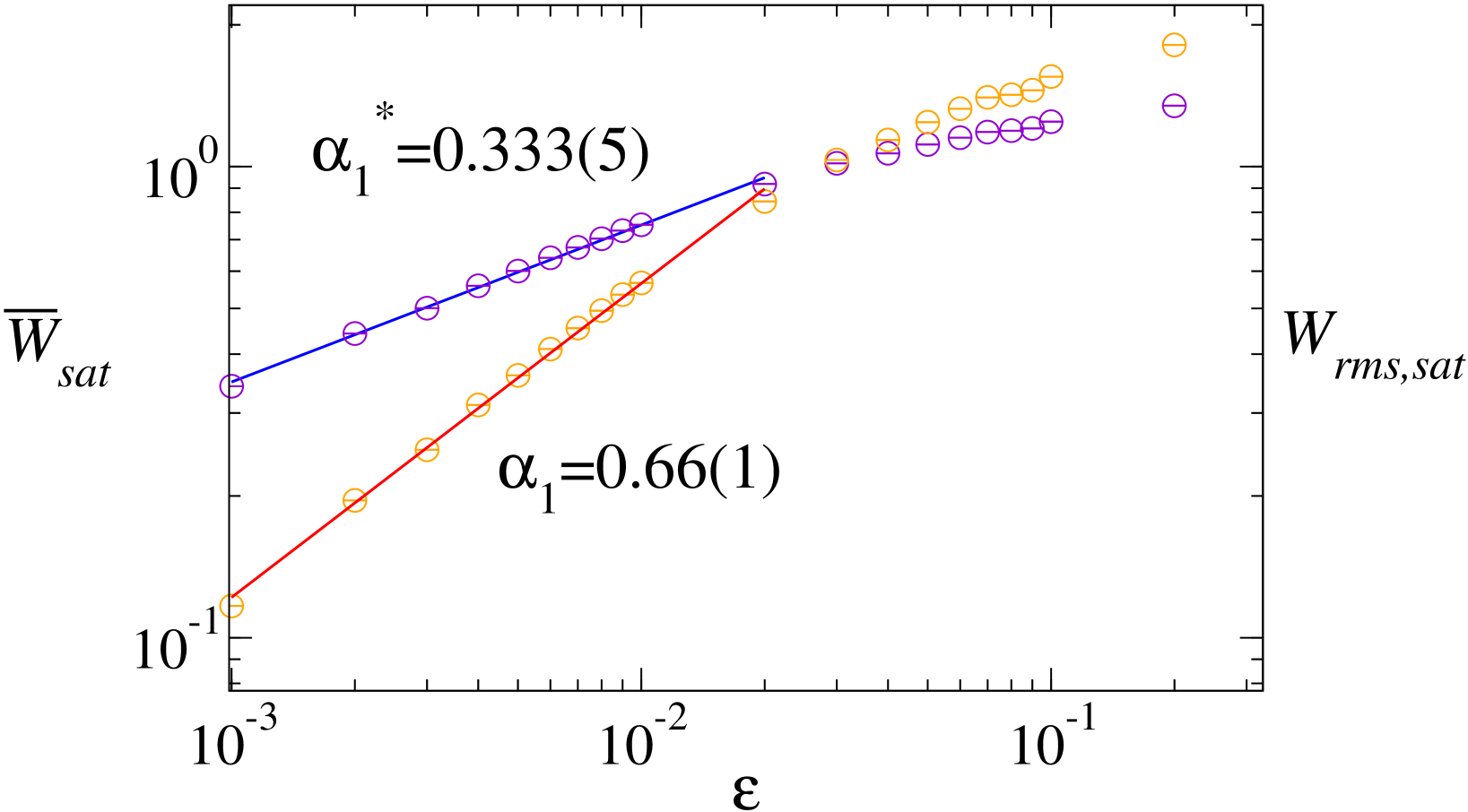}}
    
     \raggedright (b)
     \centerline{\includegraphics[width=\linewidth]{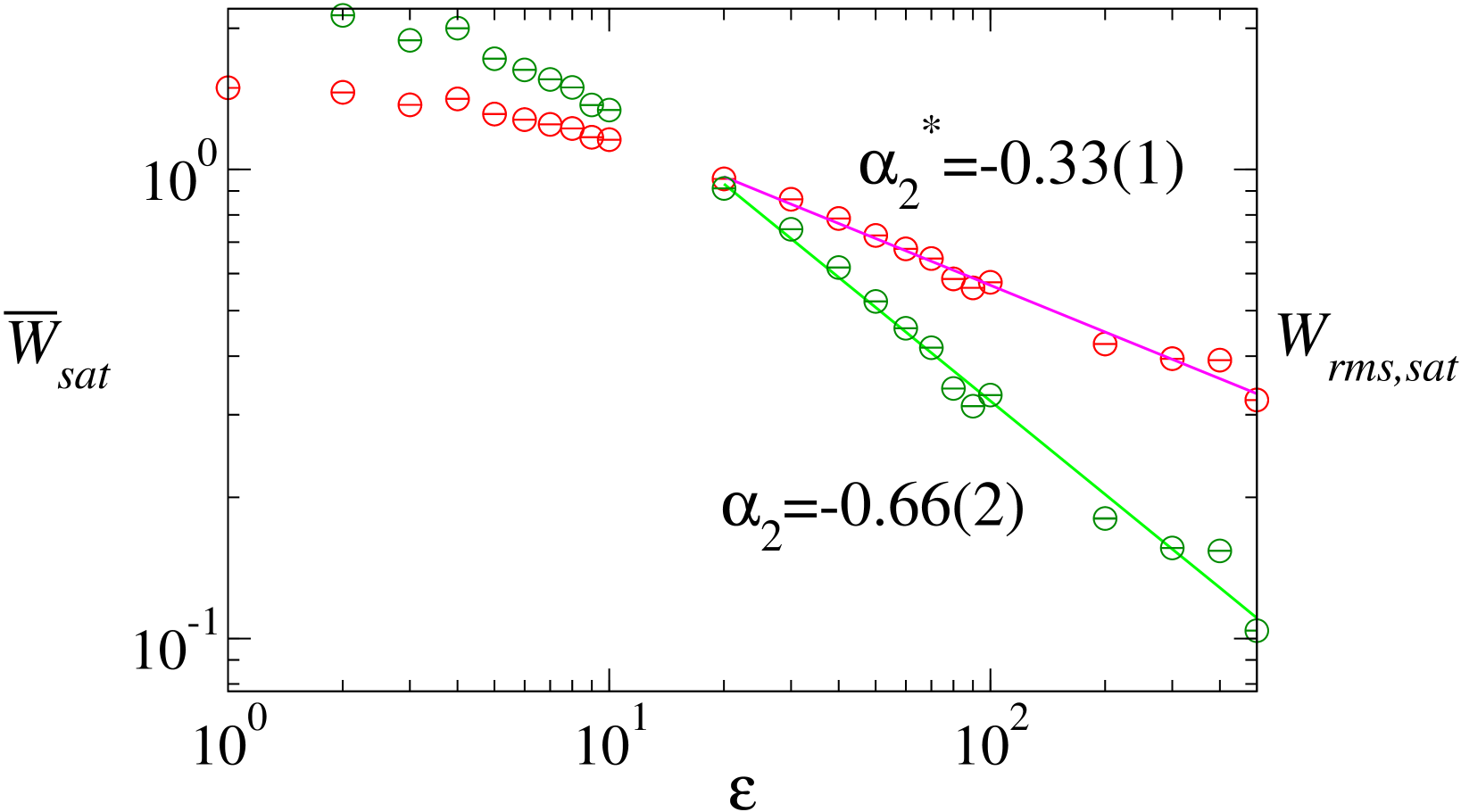}}  
    \caption{Saturated envelope measures $\overline{W}_{\mathrm{sat}}$ and $W_{\mathrm{rms,sat}}=\sqrt{\overline{W}_{\mathrm{sat}}}$ near (a) $\varepsilon\to0$ and (b) $\varepsilon\to\infty$. The plotted error bars characterize fluctuations of the accumulated estimates across the last $100$ trajectories, not independent-ensemble confidence intervals. The exponent of the square-root quantity is constrained to be half that of $\overline W$ by construction.}
    \label{fig10}
\end{figure}

For comparison on an angular scale, we also define $W_{\mathrm{rms}}=\sqrt{\bar W}$, retaining the notation used in Fig.~\ref{fig10}. It shows both definitions near the two singular limits. Deviations outside the fitted asymptotic windows indicate that a single power law does not describe the full parameter range.

We write $\overline W_{\mathrm{sat}}\propto\varepsilon^{\alpha_i}$ and $W_{\mathrm{rms,sat}}\propto\varepsilon^{\alpha_i^*}$, with $i=1$ and $2$ denoting the limits $\varepsilon\to0$ and $\varepsilon\to\infty$, respectively. The numerical fits suggest $\alpha_1\simeq0.66$ and $\alpha_2\simeq-0.66$, i.e., approximately $\pm2/3$ within the reported numerical resolution. Since $W_{\mathrm{rms,sat}}=\sqrt{\overline W_{\mathrm{sat}}}$ identically, $\alpha_i^*=\alpha_i/2$ follows algebraically, yielding approximately $\pm1/3$.

\begin{figure}[h]
    \centering
    \includegraphics[width=\linewidth]{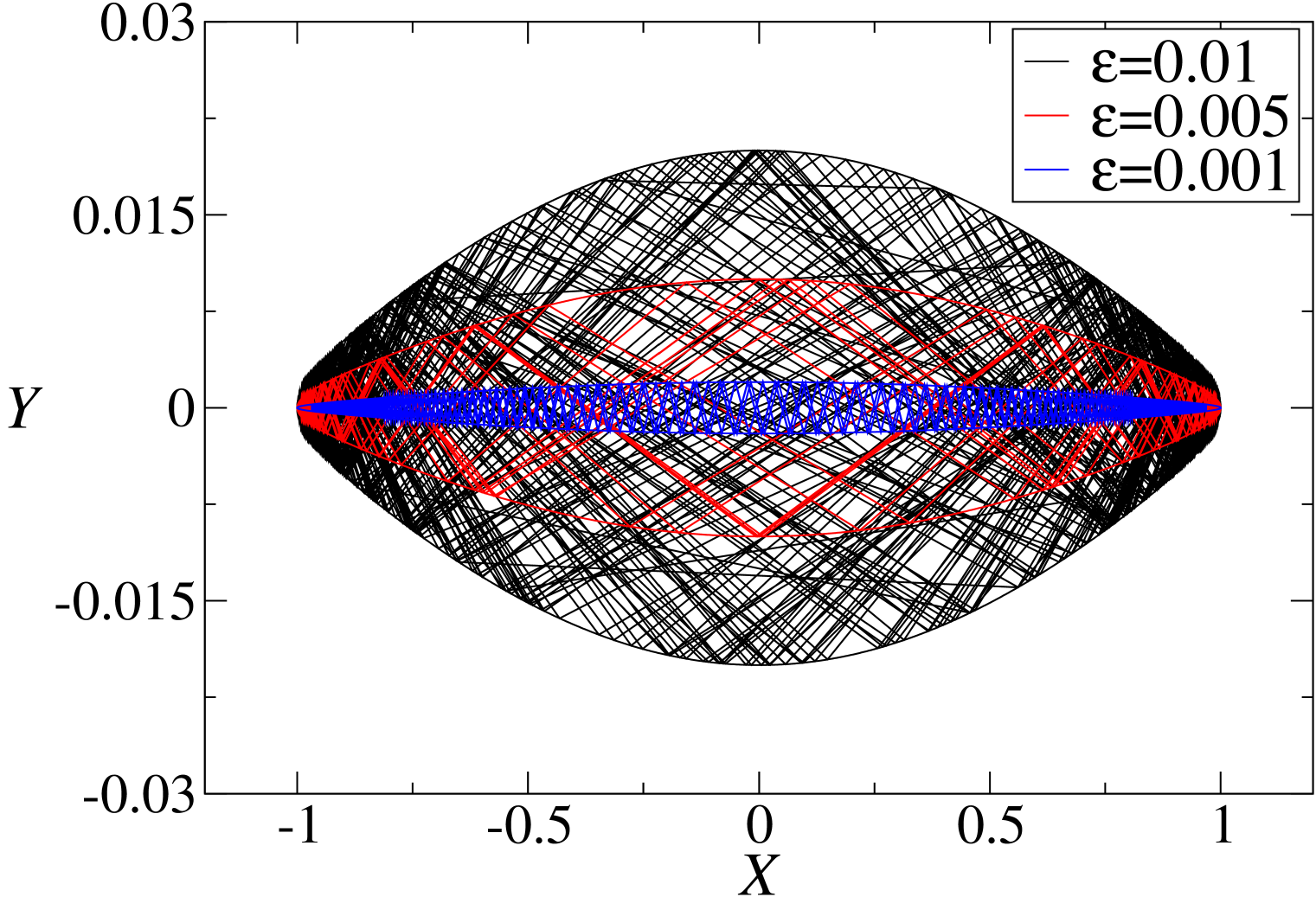}
    \caption{The first $n=2\times10^3$ collisions of a single trajectory for three values of $\varepsilon\in[10^{-3},10^{-2}]$. The plot suggests enhanced visual accumulation near the horizontal endpoints $\theta=0,\pi$; quantitative density measurements would require additional sampling.}
    \label{fig11}
\end{figure}

The measured behavior may be linked to the singular change of curvature associated with the cubic boundary profile. At the horizontal endpoints $\kappa(0)=\varepsilon^{-2}$ diverges as $\varepsilon\to0$, while at the vertical endpoints $\kappa(\pi/2)=4\varepsilon$ diverges as $\varepsilon\to\infty$. Such strongly varying curvature can affect collision statistics; studies of true cusp billiards illustrate how geometric trapping can alter correlations \cite{decay}, although the present boundary remains smooth for every finite $\varepsilon$. Figure~\ref{fig11} shows sampled trajectories near the horizontal endpoints for small $\varepsilon$. A normalized collision-density histogram would be needed to turn the visual accumulation into a quantitative transport claim.

In weakly perturbed Hamiltonian mappings, resonance widths commonly display square-root scaling and sticky motion near partial transport barriers \cite{ZASLAVSKY2002461}. The envelope measured here instead approaches two singular geometric limits, with fitted exponent magnitude approximately $2/3$ for $\overline W$. Whether that numerical exponent persists for other eye-like profiles and for coordinate-independent measures of chaotic transport is an open question; the present data support a geometry-dependent scaling law, not yet a general universality classification.

\section{Geometric response and transition-like behavior}
\label{xsec5}

The disappearance of the sampled chaotic envelope toward two singular geometric limits suggests an analogy with continuous critical behavior \cite{sethna2021statistical}. To assess the extent of this analogy, we distinguish between a geometric order-parameter-like observable and a thermodynamic order parameter, identify an explicit response function, and examine the invariant structures that affect transport.

Thermodynamic descriptions have been developed for time-dependent dissipative billiards, where boundary motion and inelastic collisions govern energy exchange and relaxation \cite{leonel2016thermodynamics}. In the present static, elastic billiard, energy is conserved, and the response function characterizes changes in the geometric extent of the sampled chaotic envelope.

For every finite $\varepsilon>0$, the boundary retains reflection symmetry across both Cartesian axes and central inversion. These discrete geometric symmetries are not spontaneously broken as $\varepsilon$ varies. What changes is the organization of the collision dynamics: rotational invariant curves, libration islands, and chaotic layers coexist, while the limit $\varepsilon\to0$ and the rescaled limit $\varepsilon\to\infty$ cease to define nondegenerate two-dimensional billiards. The axial period-two trajectories approach the normal-collision directions of the corresponding line segments. We therefore describe the phenomenon as singular geometric reorganization and loss of regular two-dimensional dynamics rather than as an established symmetry-breaking transition.

A practical order-parameter-like quantity is the effective angular envelope scale $W_{\mathrm{rms,sat}}=\sqrt{\overline W_{\mathrm{sat}}}$. It tends toward zero in the two fitted asymptotic limits. Near $\varepsilon\to0$, we write $W_{\mathrm{rms,sat}}\simeq A_-\varepsilon^{\alpha_1^*}$, where $\alpha_1^*\simeq1/3$. The associated \emph{geometric response function}, retaining the notation $\chi$, is
\begin{equation}
    \chi_{\varepsilon}=\frac{dW_{\mathrm{rms,sat}}}{d\varepsilon}\simeq A_-\alpha_1^*\varepsilon^{\alpha_1^*-1}.
\end{equation}

For $0<\alpha_1^*<1$ this derivative diverges as $\varepsilon\to0$. In the other limit, the natural small parameter is $\delta=1/\varepsilon$, proportional to the inverse vertical-to-horizontal aspect ratio. If $W_{\mathrm{rms,sat}}\simeq A_+\delta^{\alpha_+^*}$ with $\alpha_+^*\simeq1/3$, the corresponding response is $\chi_\delta=dW_{\mathrm{rms,sat}}/d\delta\simeq A_+\alpha_+^*\delta^{\alpha_+^*-1}$. The use of $\varepsilon$ near the horizontal-segment limit and $\delta$ near the rescaled vertical-segment limit compares distances to the two boundaries of the parameter family. Their geometric meaning is transparent from the aspect ratio $2\varepsilon$. Equivalently, a compact shape parameter $q=2\varepsilon/(1+2\varepsilon)$ has the endpoint distances $q$ and $1-q$, respectively. Crucially, divergence of $\chi$ is parameter dependent: for example, $dW_{\mathrm{rms,sat}}/d\varepsilon\propto\varepsilon^{-4/3}$ tends to zero as $\varepsilon\to\infty$. We therefore call $\chi$ a \emph{response function} and regard the observed endpoint singularities as critical-like scaling evidence of a phase transition.

The deformation parameter changes the local boundary tangent and therefore the scattering geometry. To characterize the local angular dependence of this scattering, consider the representative tangent-angle branch $\phi(\theta,\varepsilon)
=
\arctan\left(\frac{Y'(\theta)}{X'(\theta)}\right)$. This tangent angle enters the collision rule for $\alpha$. Substituting the derivatives gives
\begin{equation}
\phi(\theta,\varepsilon)
=
\arctan\left[-\varepsilon F(\theta)\right].
\end{equation}
where $F(\theta)=\left(1+3\sin^2\theta\right)\cot\theta$. For fixed $\theta$ away from $\sin\theta=0$, the small-$\varepsilon$ expansion of the principal arctangent gives

\begin{equation}
\phi(\theta,\varepsilon)
=
-\varepsilon F(\theta)
+\frac{\varepsilon^3}{3}F^3(\theta)
-\frac{\varepsilon^5}{5}F^5(\theta)
+O(\varepsilon^7)
\end{equation}
The leading term is $\phi(\theta,\varepsilon)\sim-\varepsilon F(\theta)$ at such fixed angles. For $\varepsilon\to\infty$, and fixed $\theta$ away from zeros of $F$, one may use the identity $\arctan(-u)
=
-\operatorname{sgn}(u)\frac{\pi}{2}
+
\arctan\left(\frac{1}{u}\right)$ in the preceding tangent-angle expression. Expanding the last arctangent gives

\begin{equation}
\phi(\theta,\varepsilon)
=
-\operatorname{sgn}[F(\theta)]\frac{\pi}{2}
+
\frac{1}{\varepsilon F(\theta)}
-
\frac{1}{3\varepsilon^3F^3(\theta)}
+
...
\end{equation}
The algebraic manipulation of $F(\theta)$ finally leads to
\begin{equation}
\phi(\theta,\varepsilon)
\sim
-\operatorname{sgn}(\cot\theta)\frac{\pi}{2}
+
\frac{1}{\varepsilon}
\frac{\tan\theta}{1+3\sin^2\theta}
\end{equation}
The correction to the limiting tangent angle is therefore of order $\varepsilon^{-1}$ at fixed $\theta$ where $F(\theta)\ne0$. These expansions are not uniform near the angular points where $F$ vanishes or diverges, precisely the regions requiring care in the singular geometric limits.

Finally, regular islands and their surrounding partial barriers govern transport through stickiness: a chaotic trajectory may spend long intervals close to an island boundary before escaping to another region of the chaotic component \cite{altmann2013}. Such structures are important dynamical obstacles. The coexistence of invariant caustics, elliptic periodic orbits, and sticky chaotic layers provides a concrete phase-space mechanism for the geometry-dependent statistics.

\section{Summary and conclusions}
\label{xsec6}

We have shown how boundary geometry organizes the coexistence of regular and chaotic motion in a static eye-shaped billiard approaching two singular geometric limits, $\varepsilon\to0$ and $\varepsilon\to\infty$ (the latter after uniform spatial rescaling). For every finite positive $\varepsilon$, the boundary remains smooth and strictly convex, supporting a nondegenerate conservative billiard. The sampled phase spaces contain elliptic islands, chaotic components, and invariant curves. Orbital instability varies nonmonotonically with the deformation: the average maximal Lyapunov exponent sampled in the chaotic component reaches its largest reported value near $\varepsilon=0.5$, with $\bar{\lambda}_{\mathrm{sat}}\simeq0.491(5)$.

Our analytical results identify the periodic-orbit structure underlying this geometric reorganization. The principal period-two diameters undergo stability changes at $\varepsilon=1/(2\sqrt2)$ and $\varepsilon=1$, and two diagonal period-two orbits exist only between these values. The noncanonical Jacobian characterizes coordinate-area deformation while preserving the Birkhoff measure. Smoothness and strictly positive curvature also ensure the existence of Lazutkin invariant caustics near grazing incidence for every finite positive $\varepsilon$. These invariant curves impose persistent barriers to chaotic transport, even as the interior phase space undergoes substantial changes.

The principal numerical finding is a common scaling pattern at the two singular endpoints. The mean squared-angle envelope width of the sampled chaotic component decreases toward zero with fitted exponents approximately $+2/3$ and $-2/3$, respectively. Their nearly equal magnitudes suggest a shared scaling structure despite the different limiting geometries. The corresponding effective angular scale, defined as the square root of this width, consequently has exponents approximately $+1/3$ and $-1/3$. This relation follows directly from the observable's definition.

The scaling laws imply a divergent geometric response $\chi$ when the effective angular scale is differentiated with respect to the chosen distance from either singular limit. This endpoint behavior motivates an analogy with continuous critical phenomena, while the system retains its reflection symmetries for every finite $\varepsilon$. The analogy is grounded in the measured envelope scaling and its associated response, together with the geometric reorganization of periodic orbits and invariant barriers.

These results identify a route to the suppression of the sampled chaotic envelope through singular geometric deformation, accompanied by the persistence of near-boundary regular motion at every finite parameter. The eye-shaped billiard thus provides a tractable setting in which analytical stability conditions, invariant transport barriers, and numerical scaling laws can be examined within a single conservative system. Deriving the observed $2/3$ exponent from the collision dynamics and testing its persistence across other boundary families are the next steps toward determining the generality of this scaling behavior.

\section*{Acknowledgements}
A.K.P.F. acknowledges CAPES (No.~88887.990665/2024-00), the Fulbright Program and the Fulbright Commission in Brazil -- Fulbright/CAPES Doctoral Dissertation Research Award (Process 2026--2027) for financial support. E.D.L. acknowledges support from CNPq (304398/2023-3) and FAPESP (2025/14544-0).

\section*{Author Contributions}
\textbf{A.K.P.F.:} Conceptualization, Validation, Formal analysis, Investigation, Visualization, Writing – original draft. \textbf{E.D.L.:} Methodology, Investigation, Writing – review \& editing, Supervision, Project administration. \textbf{D.F.M.O.:} Methodology, Investigation, Writing – review \& editing, Supervision. 
\bibliography{paper_V1}

\end{document}